\documentclass[preprint,10pt,floatfix,nofootinbib]{revtex4}
\usepackage{graphicx}
\begin{document}

%\input TD-comments
%\oddsidemargin  0.00in
%\evensidemargin 0.00in

%\doublespace
%\tighten

\title{{\it In situ} radioglaciological measurements near Taylor Dome,
Antarctica and implications for Ultra-High
Energy (UHE) neutrino astronomy\\}

%GIVE ALL PARAMETRIZATIONS OF N(Z).

\author{
D.~Z.~Besson$^1$,
J.~Jenkins$^2$,
S.~Matsuno$^3$,
J.~Nam$^4$,
M.~Smith$^2$,
S.~W.~Barwick$^4$,
J.~J.~Beatty$^5$, 
W.~R.~Binns$^6$,
C.~Chen$^7$,
P.~Chen$^7$,
J.~M.~Clem$^8$,
A.~Connolly$^9$,
P.~F.~Dowkontt$^6$,
M.~A.~DuVernois$^{11}$, 
R.~C.~Field$^7$,
D.~Goldstein$^4$,
P.~W.~Gorham$^3$,
A.~Goodhue$^9$,
C. Hast$^7$,
C.~L.~Hebert$^3$,
S.~Hoover$^9$,
M.~H.~Israel$^6$,
J.~Kowalski$^3$,
J.~G.~Learned$^3$,
K.~M.~Liewer$^{11}$,
J.~T.~Link$^{1,12}$,
E.~Lusczek$^{10}$,
B.~Mercurio$^5$,
C.~Miki$^3$,
P.~Mio\v{c}inovi\'c$^3$,
C.~J.~Naudet$^{11}$,
J. Ng$^7$,
R.~Nichol$^5$,
K. Palladino$^5$,
K. Reil$^7$,
A.~Romero-Wolf$^3$,
M.~Rosen$^3$,
L.~Ruckman$^3$,
D.~Saltzberg$^9$,
D.~Seckel$^8$,
G.~S.~Varner$^3$,
D. Walz$^7$,
F.~Wu$^4$}
\vspace{2mm}
\noindent
\affiliation{
$^1$Dept. of Physics and Astronomy, Univ. of Kansas, Lawrence, KS 66045. $^2$U.S. Antarctic Program, McMurdo Station, Antarctica. $^3$Dept. of Physics and Astronomy, Univ. of Hawaii, Manoa, HI 96822.  $^4$Univ. of California, Irvine CA 92697.  $^5$Dept. of Physics, Ohio State Univ., Columbus, OH 43210. $^6$Dept. of Physics, Washington Univ. in St. Louis, MO 63130. $^7$Stanford Linear Accelerator Center, Menlo Park, CA, 94025. $^8$University of Delaware, Newark, DE 19716. $^9$Dept. of Physics and Astronomy, Univ. of California, Los Angeles, CA 90095. $^{10}$School of Physics and Astronomy, Univ. of Minnesota, Minneapolis, MN 55455. $^{11}$Jet Propulsion Laboratory, Pasadena, CA 91109. $^{12}$Currently at NASA Goddard Space Flight Center, Greenbelt, MD, 20771.}

\begin{abstract}
Radiowave detection of the Cherenkov radiation produced by
neutrino-ice collisions requires an understanding of the 
radiofrequency (RF) response of cold polar ice.
We herein report on a 
series of radioglaciological 
measurements performed
approximately 10 km north of Taylor Dome
Station, Antarctica from Dec. 6, 2006 -- Dec. 16, 2006.
Using RF signals broadcast from: a) an
englacial discone, submerged to a depth of 100 meters and broadcasting
to a surface dual-polarization horn receiver,
and b) a dual-polarization horn 
antenna on the surface transmitting signals which reflect off the
underlying bed and back up to the surface receiver, we have made 
time-domain estimates of both the real (index-of-refraction ``$n$'')
and imaginary (attenuation length ``$L_{atten}$'') 
components 
of the complex ice dielectric constant. We have also measured the uniformity
of ice response along two orthogonal axes in the horizontal plane.
We observe an apparent 
wavespeed asymmetry of order 0.1\% between two
orthogonal linear polarizations projected into the horizontal plane,
consistent with 
some previous measurements, but somewhat lower than others. 
\end{abstract}

\date{\today}

\maketitle

\section*{Introduction}
\label{intro.sec}
The Antarctic icecap is the world's largest stable, homogeneous
surface feature, comprising 75\% of the current freshwater reserves on
the planet. The pristine nature of the ice, with relatively few defects
or impurities, results in exceptional transmission
properties for both electromagnetic and acoustic/seismic signals. 
At wavelengths of 300 nm(/1 m), electromagnetic
attenuation lengths are of order
200 m(/2000 m)\citep{Buford1997,Bar2005}. 
%Attenuation lengths of order 10 km are expected for acoustic signals in the 10-30 kHz frequency range\cite{Price-2005}. At lower frequencies, the long seismic attenuation length facilitates trans-Earth  seismic observing at the South Pole\citep{SPRESO}. 

The lack
of substantial
human and animal activity on the continent makes the icecap
an ideal locale for experimental efforts seeking to measure rare
collisions of extra-terrestrial objects, both 
microscopic\citep{Kr2006,Bar2005} and macroscopic (meteorites,
e.g.\citep{ANSMET}) with the icecap itself.
The long scale of signal transmission allows a radio sensor
to probe an extremely large volume for englacially generated
radiowave signals. 
In the case where 
ultra-high energy cosmic-ray neutrinos are measured via the
coherent Cherenkov
radiation produced subsequent to their collision with ice
molecules, ice properties must be understood.
In particular,
the attenuation, absorption, and possible
de-polarization
of the resulting electromagnetic signals due to the
ice intervening between the interaction point and the detection
sensor must be quantified if one is to accurately reconstruct
the four-momentum of the initial neutrino.
Measurement of Antarctic ice properties over a large
footprint is particularly important for the ANITA experiment\citep{Bar2006}, 
which
seeks to register neutrino-induced radiowave signals using
a suite of high-bandwidth
horn antennas mounted on a gondola. From a height of 38 km, 
ANITA monitors a mass of ice out to the horizon 680 km away.

The electromagnetic response of some medium is typically
expressed in terms of a complex dielectric `permittivity'
$\epsilon=\epsilon'(\omega,{\hat n})+i\epsilon''(\omega,{\hat n})$,
where $\omega$ is the angular frequency and ${\hat n}$ is the
polarization vector. The real part of the dielectric
constant ($Re(\epsilon)=\epsilon'$) is related to the electromagnetic
wavespeed via: $v'=c/\sqrt{|\epsilon'|}$; 
in the limit of small $\tan\delta$,
the imaginary
part ($Im(\epsilon)=\epsilon''$) gives the power loss through the medium via:
loss (dB/m)=8.686($\omega/2c_0$)($\sqrt{\epsilon'}\tan{\delta}$)\citep{Jackson75}. %\footnote{Alternately, in terms of the conductivity $\sigma$, this expression can be written as $L=4.343\sigma\sqrt{\mu_0\over{\epsilon_0\epsilon'}}$. Equating these expressions relates measurements of the electrical conductivity $\sigma$ (in units of Seimans) to $tan{\delta}$ by $\epsilon''=\sigma/(\epsilon_0\omega)$, with $\omega$ the angular frequency. Thus, given the loss ($L$ dB/m), the ratio of amplitudes at two points ($A_0$ and $A_1$) separated by 1 m is: $A_1/A_0=10^{-L/10}=e^{-1~m/\alpha}$, or $\alpha = 10/(Lln(10))$.}
Here, $c_0$ is the electromagnetic wavespeed {\it in vacuum}, and 
$tan{\delta}=\epsilon''/\epsilon'$.
%0.09$f(MHz)\sqrt(\epsilon')tan{\delta}$. 

%google ``antarctic ice birefringence'' - lots of nice pictures and comments on surface roughness in http://www.vexcel.com/downloads/rd/marpolar/PBAND-WP100_antarctic-ice-d.pdf, as well as extensive info on BEDMAP
The response of ice as a function of polarization (``birefringence'') has
been probed with a variety of measurements\citep{Hargreaves,Hargreaves-1978,Matsuoka-biref,Matsuoka2004,Doake2002,Doake2003}. 
Asymmetries are characterized
as differences in either wavespeed or absorption along linear
(generally orthogonal) axes. %, or alternately, in terms of a left-handed circular polarization (LCP) vs. right-handed circular polarization (RCP) basis. 
Formally, the two asymmetries ($\delta_{\epsilon'}$, real and/or
$\delta_{\epsilon''}$, imaginary) are linked by the Kramers-Kr\"onig
dispersion relation -- if one is non-zero the other must be non-zero, 
as well.
In the absence of any preferred in-ice direction, one might
expect any asymmetry to be mitigated by the random-walk nature of 
the birefringence. In such a case, over a total pathlength $l$ consisting
of $N$ unit steps, each of which is
characterized by an asymmetry $b$, the average
propagation time for each polarization axis should have a Gaussian
distribution, centered at $l/c$, with width $b\sqrt{N}l/Nc$. The asymmetry
distribution would therefore be a Gaussian of width $\sigma_b=b\sqrt{2N}l/Nc$,
centered at zero.
For 1\% birefringence ($b=0.01$), $l$=1000 m, and
values of distance scale corresponding to 
typical grain sizes ($10^{-3}$ m, or N=$10^6$), 
we expect $\sigma_b\lesssim$0.1 ns. 
By comparison,
typical neutrino-induced 
signal durations measured by ANITA 
are of order 3-4 ns, so this difference is insignificant.

The impedance mismatch of air relative to ice, or ice relative to bedrock introduces non-zero reflections at that interface. The corresponding reflection and transmission coefficients, for the perpendicular and parallel components of planar incident electric field wavefronts ($r_{\bot}$, $r_{||}$, $t_{\bot}$ and $t_{||}$, respectively), are given by the standard ``Fresnel equations for dielectric media'', in terms of the angle-of-incidence ($\theta_i$) and angle-of-transmission ($\theta_t$), and the index of refraction ($n(\omega,{\hat n})=\sqrt{\epsilon'(\omega,{\hat n})}$).
%\footnote{The interested reader is referred to websites http://scienceworld.wolfram.com/physics/FresnelEquations.html, and http://hyperphysics.phy-astr.gsu.edu/hbase/phyopt/freseq.html} \begin{center} $r_{\bot}=(n_icos\theta_i-n_tcos\theta_t)/(n_icos\theta_i+n_tcos\theta_t) = -(sin(\theta_i-\theta_t))/(sin(\theta_i+\theta_t)),$ $r_{||}=(n_tcos\theta_i-n_icos\theta_t)/(n_icos\theta_t+n_tcos\theta_i) = +(tan(\theta_i-\theta_t))/(tan(\theta_i+\theta_t)),$ $t_{\bot}=2sin(\theta_t)cos(\theta_i)/sin(\theta_i+\theta_t),$ $t_{||}=2sin(\theta_t)cos(\theta_i)/(sin(\theta_i+\theta_t)cos(\theta_i-\theta_t).$ \end{center}
For ANITA, the change in dielectric from below the firn up through the surface 
interface (i.e., air-ice) largely determines the volume of ice visible to the balloon. 
For constant voltage threshold $V_{thresh}$
(and neglecting absorption), the minimum field strength observable from the balloon is given by $E_0R=V_{thresh}$, where $E_0$ is the strength of the signal at the production point, $R$ is the distance from production point to observation point and $V_{thresh}$ is the voltage threshold at the detector, typically set to be at least three times the ambient thermal noise background, or: $V_{thresh}\sim 3V^{rms}_{thermal~noise}$. For large values of $E_0$, the observable volume is (neglecting the 38 km altitude of the balloon, and taking $R$ to be measured in the horizontal plane) disk-like. In this case, most of the sensitive volume is at the edge of the horizon and the angles of incidence relative to the balloon are close to the critical angle.

\subsection*{Scope of Experimental Work}
Our primary goal was to perform measurements important to understanding
radiowave propagation through the ice, refracted at the surface, and 
ultimately received by the ANITA gondola. To that end, a series of
studies were conducted, as summarized in Table \ref{t:intro2studies}.
\begin{table}[htpb]
\begin{tabular}{c|c|c|c}
Measurement & Transmitter$\to$Receiver & Geometry & Implications \\ 
            & Configuration            &          &              \\\hline \hline
$L_{atten}$ ($Im(\epsilon)$) & Surface horn$\to$surface horn & ``V''-axis$\equiv$14.8 degrees
	& Range of observable \\
  & reflected off bedrock      & E of true N; in horizontal plane & neutrino interactions \\ \hline
Birefringence & (same) & (same) & ANITA trigger efficiency \\ \hline \hline
$n(z)$ ($Re(\epsilon)$) & in-ice discone$\to$surface horn & ``V''-axis$\equiv$vertical c-axis &
$\theta_{critical}$, \\
$\equiv$Index-of-refraction as f(depth)	& & & ray-tracing through ice/firn \\ \hline
Surface Roughness & (same) & (same)
& Signal amplitude at gondola \\ \hline
\end{tabular}
\caption{Scope of radioglaciological measurements carried out at Taylor Dome.
The ``V'' axis defines antenna orientation, and is detailed elsewhere in
the text. \label{t:intro2studies}}
\end{table}

\section*{Experimental Technique}

\subsection*{Geometry}
\begin{figure} \centerline{\includegraphics[width=12cm]{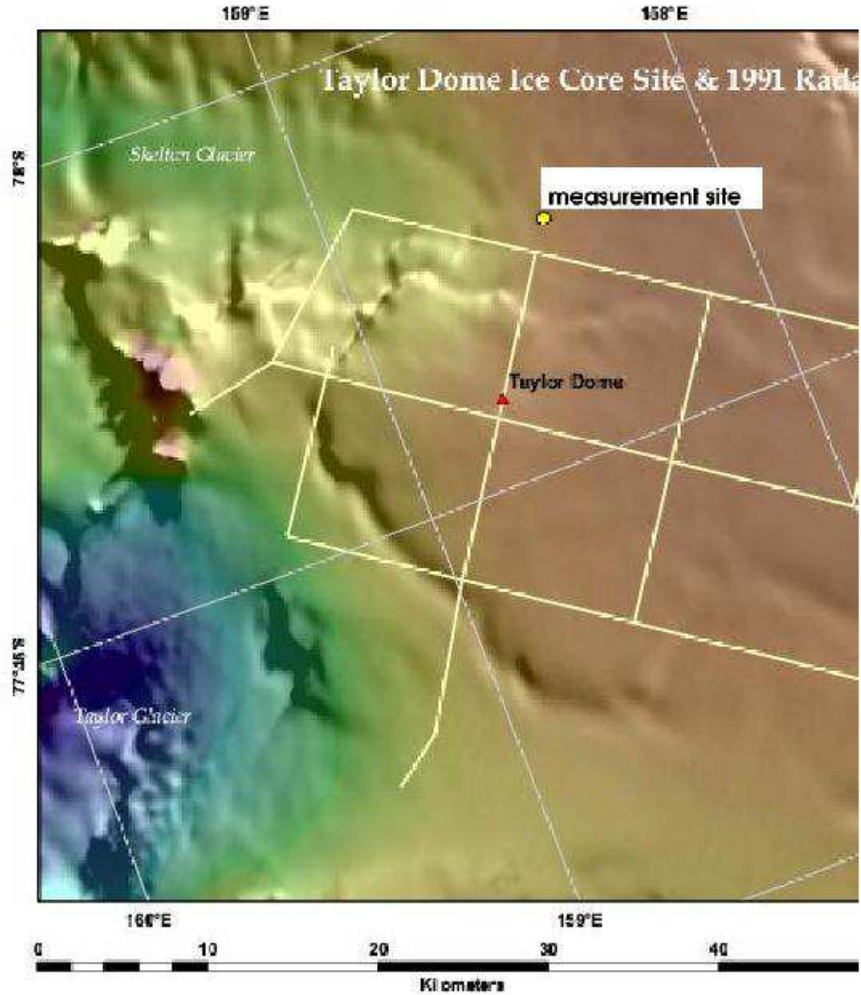}} \caption{Coordinates of measurement site (yellow circle) and main Taylor Dome base (red triangle)} \label{fig:TaylorDomeMap0.eps} 
\end{figure}

\begin{figure} \centerline{\includegraphics[width=16cm]{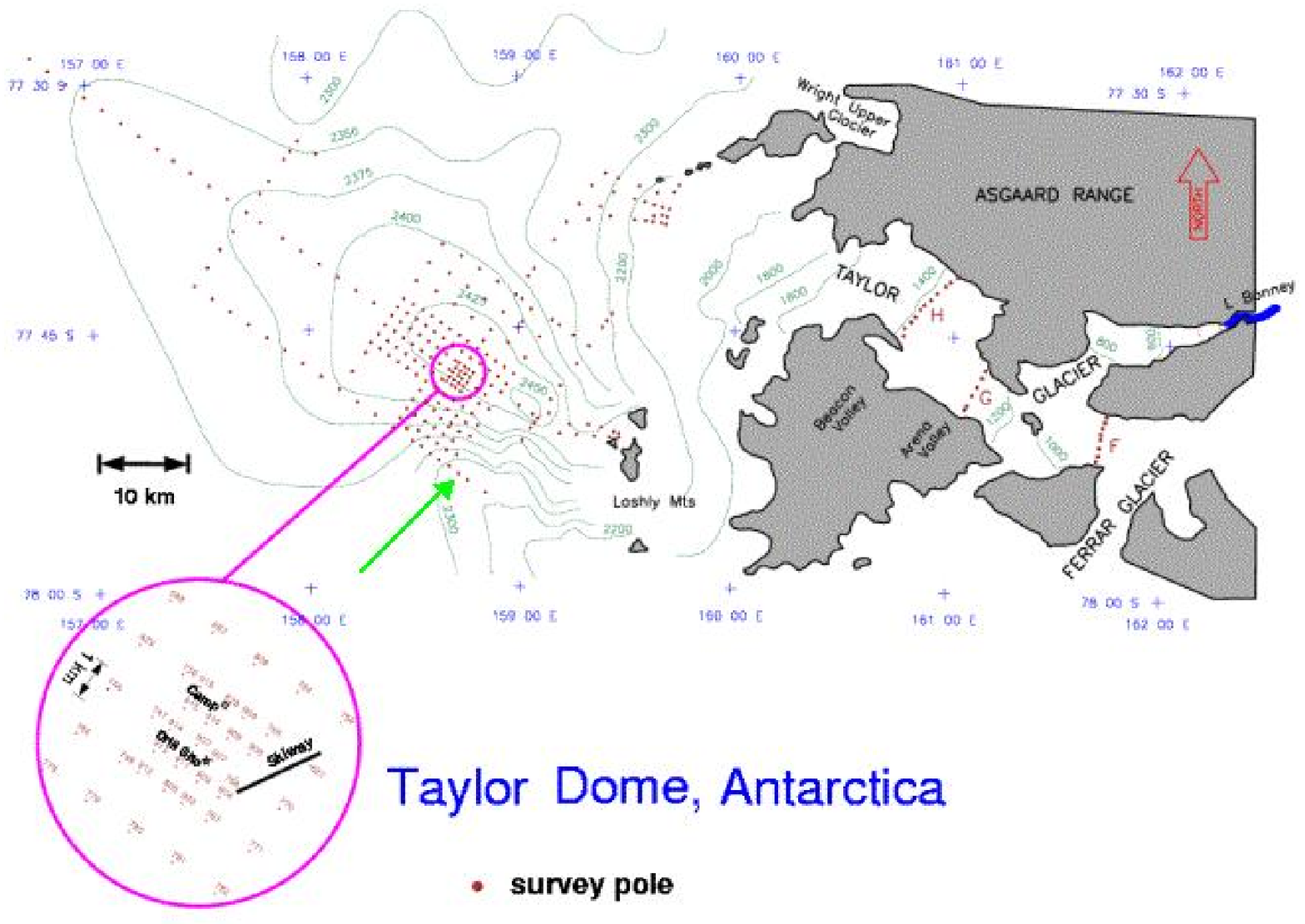}} \caption{Topographical map of experimental site. Measurements were made at location corresponding to tip of green arrow.}\label{fig:TaylorDomeMap1.eps} 
\end{figure}

The geography of the NASA Core Site 
camp is presented in Figures
\ref{fig:TaylorDomeMap0.eps} and \ref{fig:TaylorDomeMap1.eps} (tip of
green arrow). As indicated in Table \ref{t:intro2studies},
the orientation of the antennas is denoted by the two orthogonal linear
polarization `V' and `H' axes. If
both transmitter (``Tx'') and receiver (``Rx'') are aligned with
`V', the orientation is therefore
described as `VV', e.g. In the
other case of the 
vertically-oriented
in-ice discone broadcasting to a surface receiver,
only the polarization of the receiver horn is indicated (``Hpol'' or
``Vpol'', e.g.).
For the in-ice measurements described below, the VV-axis
is approximately 14.8 degrees East of true North, 
and points 
roughly in the direction of the primary Taylor Dome base. Over
a distance scale of $\sim$10~km, the VV axis approximately
coincides with the surface elevation gradient over that scale.
Over a shorter distance scale 
($\sim$0.5 km), the HH-axis coincides
roughly with the more local surface gradient.

\subsection*{Transmitted Signal Characteristics}
\begin{figure} \centerline{\includegraphics[width=12cm]{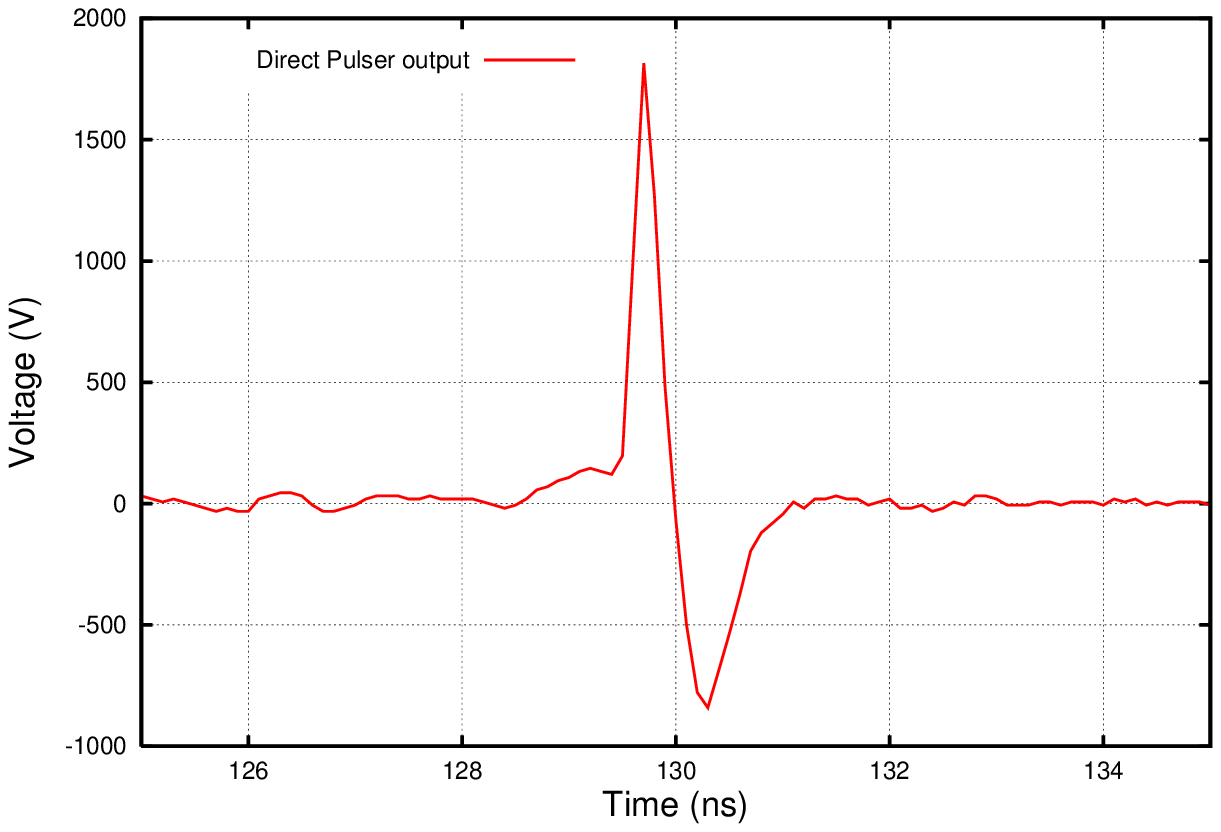}} \caption{Pulse generator output signal. Horizontal time offset from zero is arbitrary.}\label{fig:TD-FID-output-direct.eps} 
\end{figure}

\begin{figure} \centerline{\includegraphics[width=12cm]{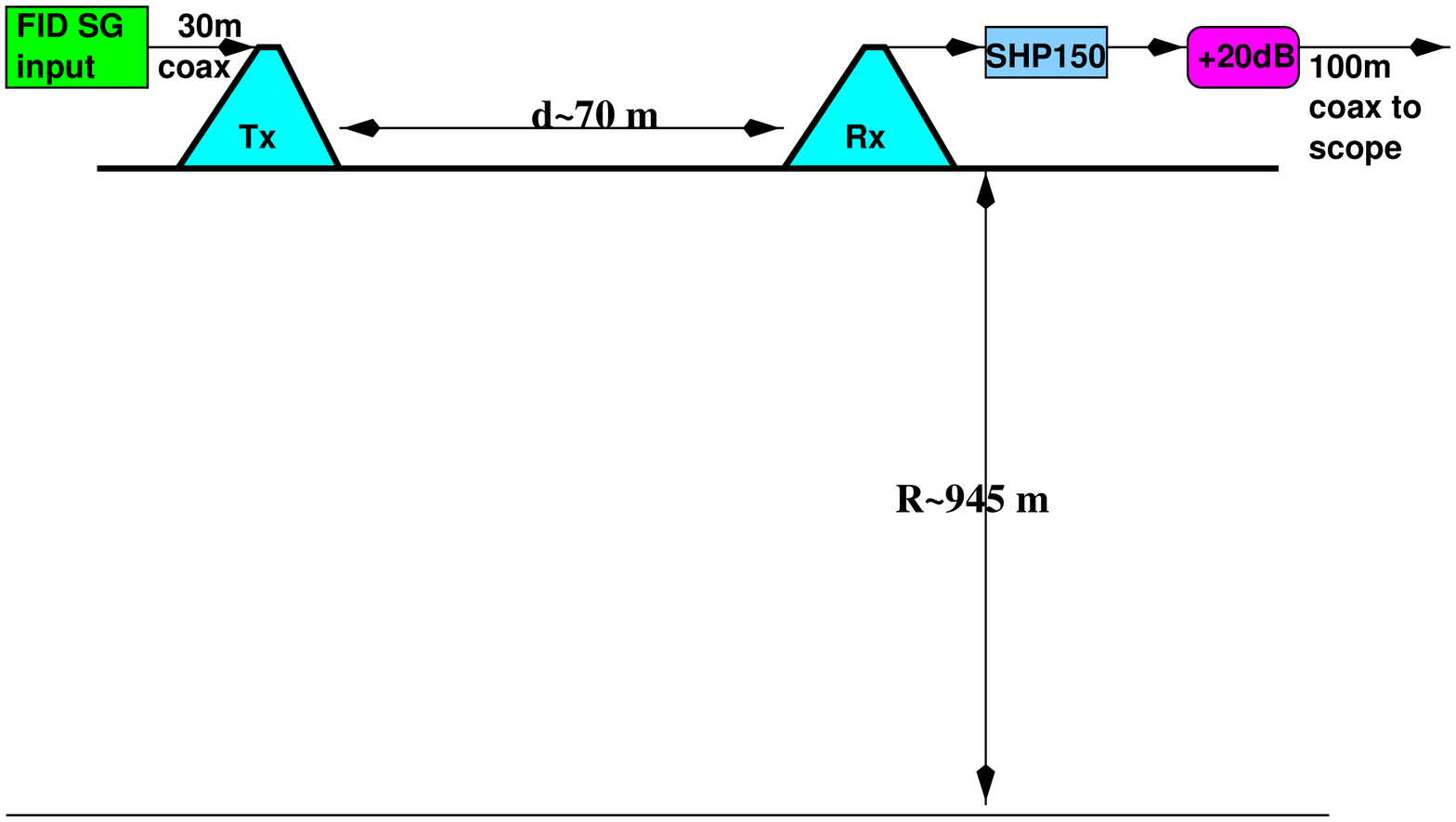}} \caption{Set-up used for experimental measurement of bottom reflection. For
in-air transmission (``${\rm S}_{12}$(air)''), horns face each other on the
surface. Bottom reflection
measurements were also made with the receiver horn
antenna displaced $\sim$100 m to the left of the transmitter,
as described later in this document.}\label{fig:TD-S11-S12.xfig.eps} 
\end{figure}

Signals are produced using a high-power (2.5 kV peak-to-peak voltage) fast
pulse generator; 
the pulser output signal is shown in Figure
\ref{fig:TD-FID-output-direct.eps}.
A short 30-meter, low-loss 0.5-inch thick 
coaxial cable connects the pulser
output to a Seavey quad ridge dual-polarization transmitting 
horn antenna (Seavey Engineering model QRG-218A), %http://www.seaveyantenna.com/PDF-RF/058-061.pdf
with separate connections along the orthogonal
V- and H- axes. 
Isolation between the two
polarizations is typically $\ge$14 dB (power). This antenna offers
excellent response in the range 200 MHz--1.5 GHz, and is identical to
those mounted on the ANITA gondola. The in-air
average boresight gain has been 
measured to be 10 dBi, with a full beamwidth of $\sim$45 degrees.
For the (effective)
attenuation length ($L_{atten}$)
measurements described below,
signals were directed either
downwards and reflected off the underlying bedrock to an identical
receiver
surface antenna, or parallel to the ice surface, with
both antennas facing each other through-air, to provide
a normalization for the in-ice reflected signal. 
(It is important to emphasize here that we measure the
average, or ``effective'' attenuation
length averaged over ice of varying temperature, whereas the
attenuation length relevant to neutrino detection is primarily that of
cold [$T\sim-$50 C] polar ice.)
For reflection off the
bedrock, the receiver horn signal was first high-pass
(Minicircuits model SHP-150) filtered
and then immediately 
amplified by 20 dB in power to
enhance the signal voltage (Figure \ref{fig:TD-S11-S12.xfig.eps}). 
In the configuration where the
surface receiver measures signals broadcast from an in-ice
discone (designed and constructed
at the University of Hawaii), 
the surface horn is inclined downwards at a cant angle of 11
degrees below the horizontal, unless otherwise specified.
The discone was lowered into
a $\sim$12 cm diameter hole, drilled 
to a depth of 100 m
in November, 2006, 
by the 2006-07 ITASE drilling team.

Following reception
at the receiver horn, signal is
conveyed through $\sim$100 m of low-loss coaxial cable to a
3 GHz-bandwidth Tektronix TDS694 digital oscilloscope, which performs
waveform capture at 2.5 GSa/sec. 
For most of the measurements described herein, 
we average over multiple samples ($\sim$60-2000, depending on the 
measurement). The single-shot jitter is observed to be approximately
200 ps in time, and 5\% in peak voltage. 

\section*{Procedure}
\subsection*{Measurement of Attenuation Length}
Attenuation of radiofrequency signals is impurity
and temperature-dependent, and shows large
variations across the Antarctic continent. A recent 
frequency-domain RF
attenuation length measurement at SIPLE
Dome in West Antarctica, for
example, gave an average field attenuation
length value of $L_{atten}\sim$238--334 m\citep{SIPLE-attenlen} in
the range $\sim$5 MHz, 
considerably
smaller than a time-domain measurement at South Pole\citep{Bar2005}. 
This lower average attenuation length is not entirely unexpected,
given the relatively warm ice temperature profile in West
Antarctica and the proximity to the ice sheet at the former location.
%The higher impurity levels expected closer to the periphery lead to an expected 1/f dependence of the attenuation length.
The ANITA sensitivity is dominated by the interior ice.
%http://igloo.gsfc.nasa.gov/wais/pastmeetings/PPT05/Posters/Price.pdf

\subsection*{Received Power Magnitude and $L_{atten}$ Determination}
Two complementary calculations permit an estimate of the average 
ice radiofrequency attenuation length at our site.
First, we compare the signal
amplitude $V_{ice}$ measured through the in-ice path (${\rm S}_{12}$(ice)) 
normalized relative to the signal
amplitude $V_{air}$ measured when the transmitting horn broadcasts 
along boresight to the
receiver horn in-air (${\rm S}_{12}$(air)). 
Knowing the distance
between the horns in-air ($d_{air}$) and the
round-trip distance of the signal path through ice $d_{ice}$, 
%considering the air to be non-absorptive,
and attributing 
all losses greater than
(assumed spherical) 1/r amplitude spreading to ice
attenuation, we can
use $V_{ice}/V_{air}=(G_{ice}/G_{air})\times(d_{air}/d_{ice})e^{-d_{ice}/\langle L_{atten}\rangle}$
to extract the mean field attenuation length $\langle L_{atten}\rangle$.
Implicit in this expression is the assumption that the voltage
decreases as 1/r; i.e., that the reflection off the bedrock is
coherent. We consider the validity of this 
assumption later. 
Here, $G_{air}$ and $G_{ice}$ are the gains for transmission
through-air and through-ice, respectively. Since the 
beamwidth decreases, directivity, or forward gain, increases
in the higher-dielectric medium.
In our case, assuming the antennas are horizontal on the surface
to within 0.1 radians, and the
antennas ``see'' only ice, $G_{air}/G_{ice}\sim$ 10 dBi/15 dBi.
%(A more realistic estimate might an ice-air average gain.)

%\subsubsection*{Extraction of attenuation length from Friis Equation}
Second, rather than normalizing to the in-air signal,
we can ``dead reckon'' the expected signal at the antenna using
the Friis equation.
Knowing the total cable length
($\sim$125 m), the cable loss per unit
length ($\sim$5 dB/100 m at 500 MHz), the maximum signal
amplitude at the output of the pulser (2.5 kV), the measured
reflected signal amplitude $V_{ice}$, the bandpass of the antennas
($\sim$1 GHz), the net gain of the amplifiers + filters
in the system (+$\sim$18 dB), and the forward in-ice
gain of the horn antennas, 
we can determine ``absolutely'' the   
average attenuation length.

\subsection*{${\rm S}_{12}$(air) measurements}
\begin{figure}[h]
\begin{minipage}{18pc}
\includegraphics[width=8cm]{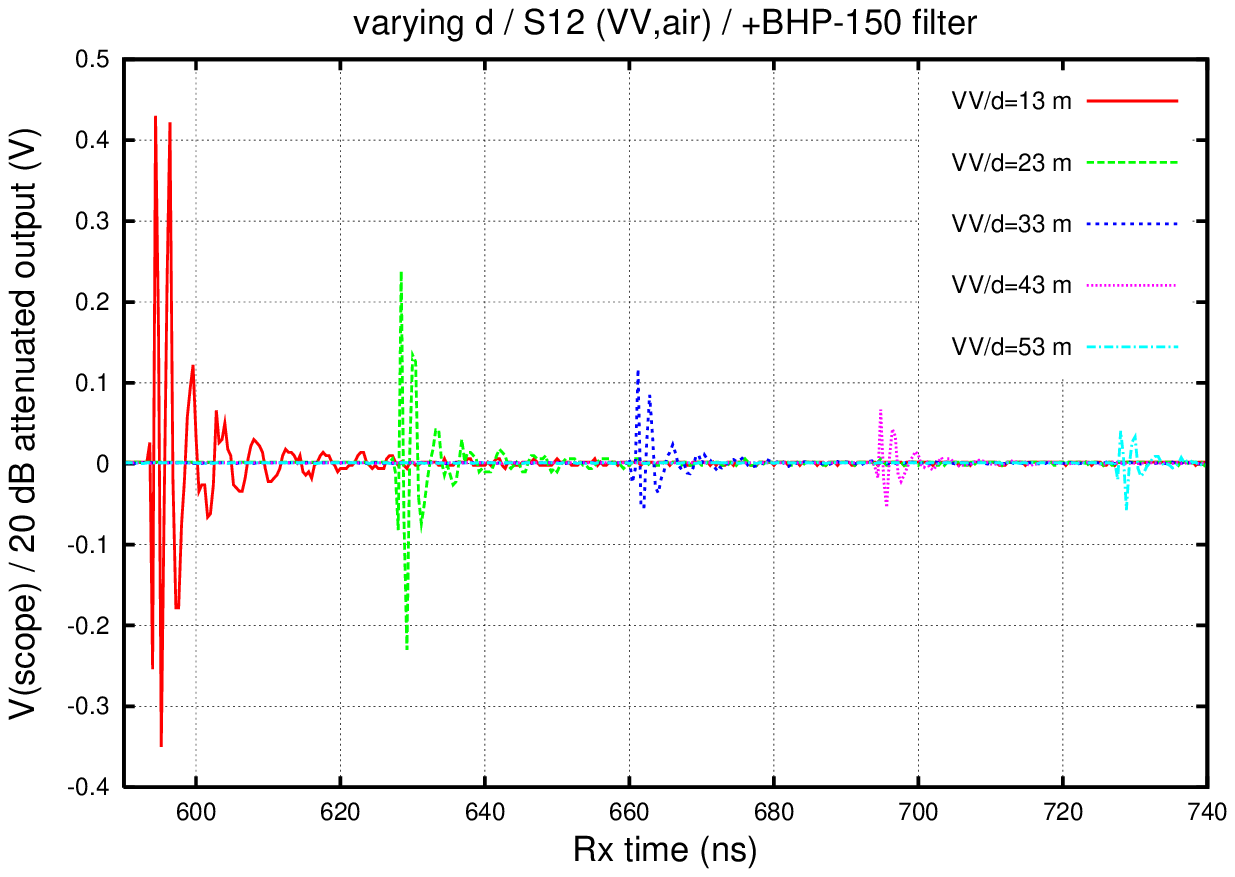}
\caption{VV signal observed for in-air transmission,
as a function of separation between transmitter and antenna. Overlaid
is the expectation that, in the absence of interference effects, the
amplitude should vary as 1/r. Note that the extracted propagation 
velocity agrees with $c$ to within 2\% (50 m/134 ns).} 
\label{fig:TD-11dec-S12-air-variable-d-VV.eps} 
\end{minipage}
\hspace{0pc}
\begin{minipage}{18pc}
\includegraphics[width=8cm]{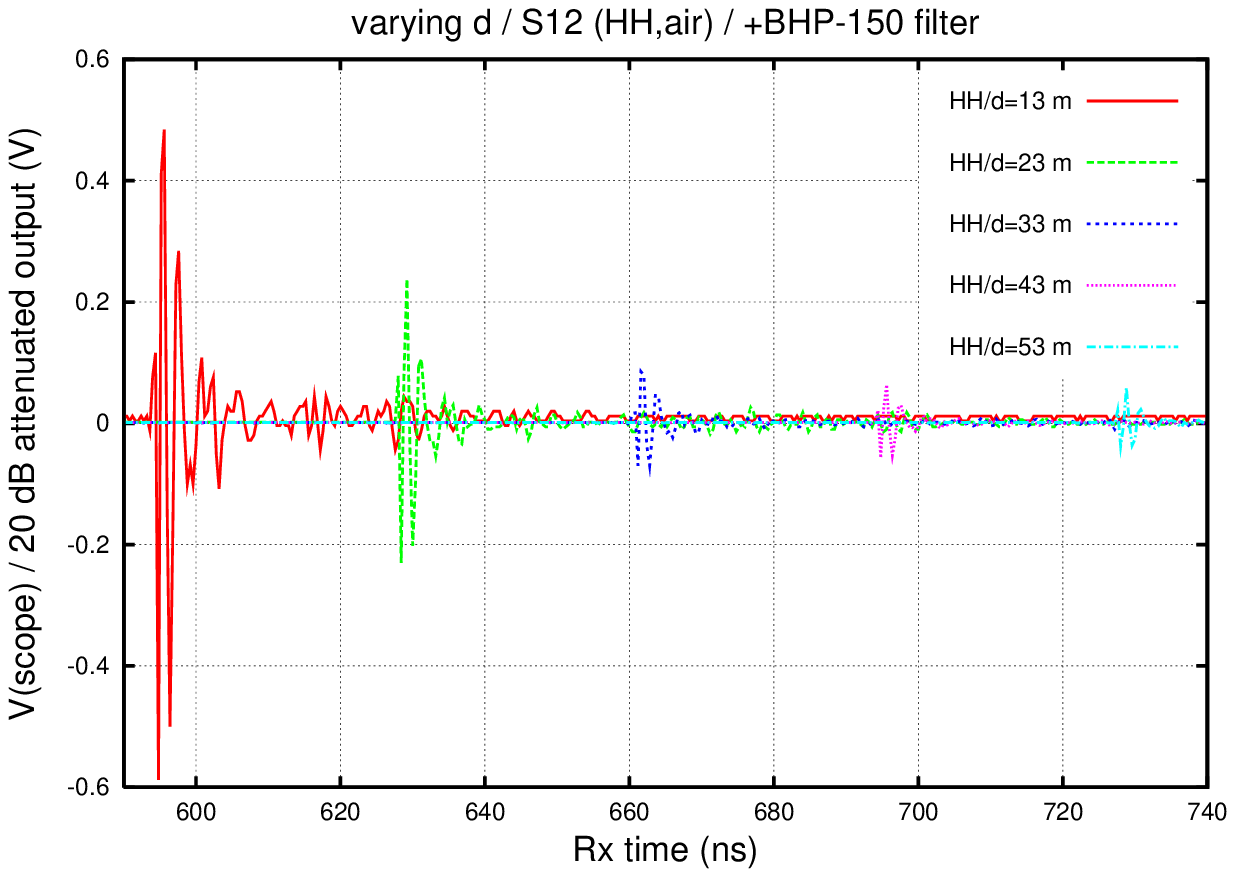}
\caption{HH signal observed for in-air transmission,
as a function of separation between transmitter and antenna. Relative 
transmitter and receiver geometry identical to
that for the previous Figure; only
the broadcast polarization has been changed.}
\label{fig:TD-11dec-S12-air-variable-d-HH.eps}
\end{minipage} 
\end{figure}

For the in-air path, signals were broadcast over a variety of
separation distances. To ensure that the trigger signal
does not saturate the amplifier, the output
of the signal generator was attenuated by 20 dB (power);
by contrast, for broadcasts through ice, the signal from the
receiver antenna was amplified by 20 dB (power). The raw
received signals
shown for HH and VV are comparable (Figures \ref{fig:TD-11dec-S12-air-variable-d-VV.eps}
and \ref{fig:TD-11dec-S12-air-variable-d-HH.eps}; in these and
subsequent figures, unless explicitly indicated, waveforms have not
been corrected for cable attenuation.). 
By contrast, 
cross-talk (VH or HV) is
observed to be $\lesssim$10\% in voltage amplitude (Figure 7).
\message{CHECK FIGURE NUMBERS HERE - SOME ARE HARDWIRED}
%(Figure \ref{fig:TD-11dec-S12-air-variable-d-VH.eps}). 
We note that the amplitudes for HH and VV are quite comparable
%(Figure \ref{fig:TD-11dec-S12-air-variable-d-VH.eps}).
(Figure 8). Overlaid on the VV- and HH- datasets is the
expectation that the signal amplitude should decrease as 1/r for
3-dimensional spreading of spherical wavefronts. We observe
deviations of $\sim$25\% from this naive expectation, which
translates into an uncertainty of order 10\% in our extracted
field attenuation lengths. Although the far-field approximation
should be valid for all these data, the closeness of the antennas to 
the ice-air boundary and multipath interference effects may 
complicate interpretation of the in-air data. We consider 
possible interference effects
now.

\message{IS THERE VH DATA FOR 13 METERS?}
\begin{figure}[h]
\begin{minipage}{18pc}
\includegraphics[width=8cm]{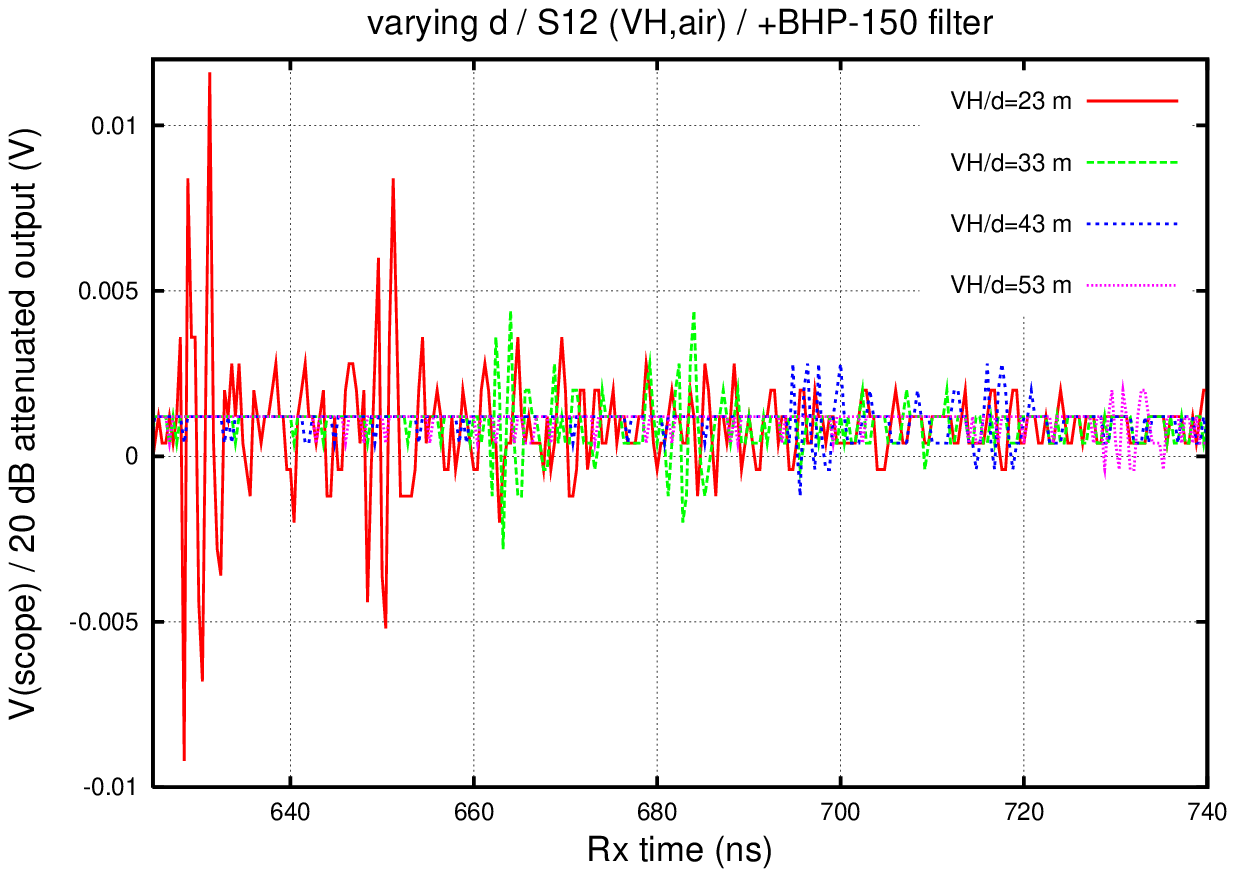} 
\label{fig:TD-11dec-S12-air-variable-d-VH.eps} 
\caption{HV signal observed for in-air transmission,
as a function of separation between transmitter and antenna.} 
\end{minipage}
\hspace{0pc}
\begin{minipage}{18pc}
\includegraphics[width=8cm]{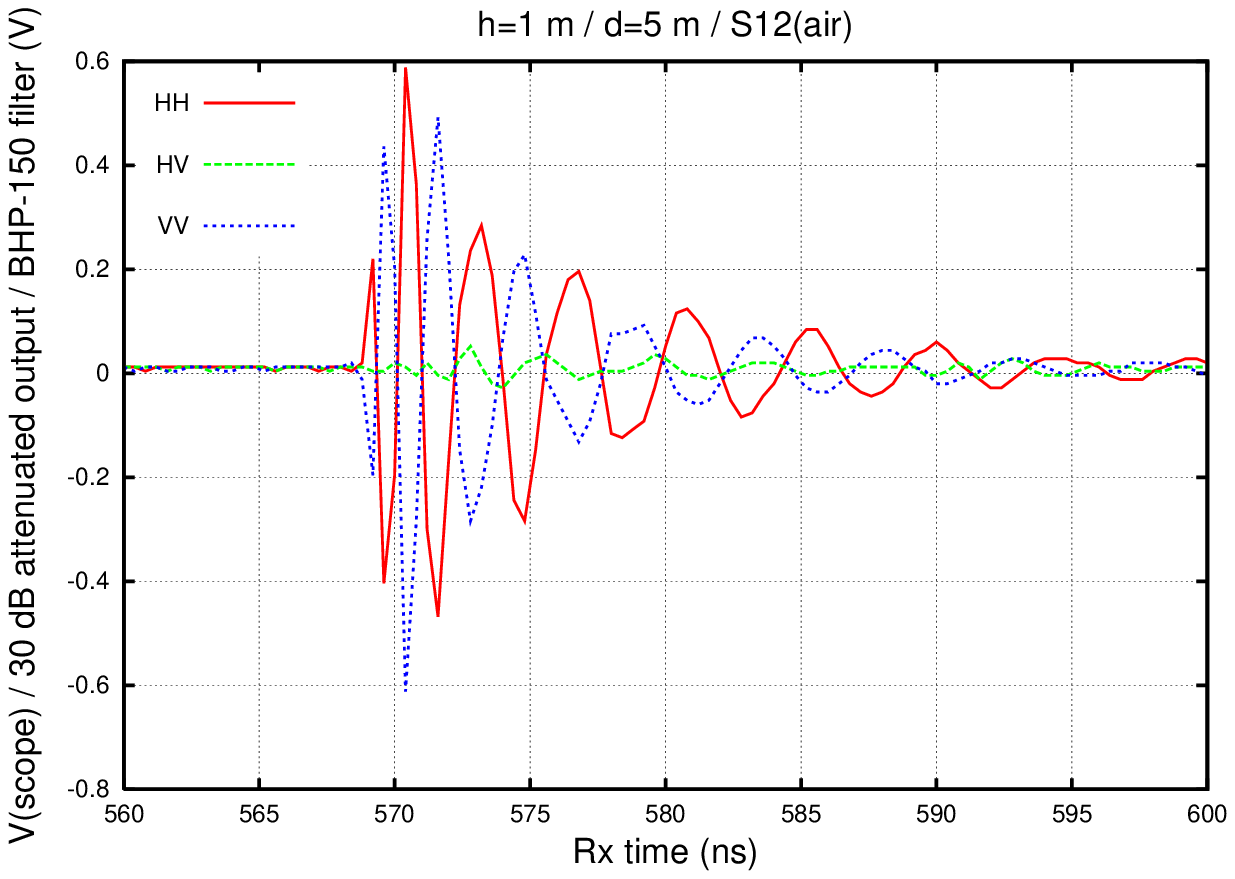} \caption{Comparison of
in-air signal strength observed for VV and HH with both Tx and Rx elevated approximately 1 m above snow surface.} \label{fig:TD0.eps} 
\end{minipage} 
\end{figure}
%Our default configuration employed the 150 MHz high-pass filter. 

\subsubsection*{Interference considerations for through-air broadcasting.}
\begin{figure} \centerline{\includegraphics[width=12cm]{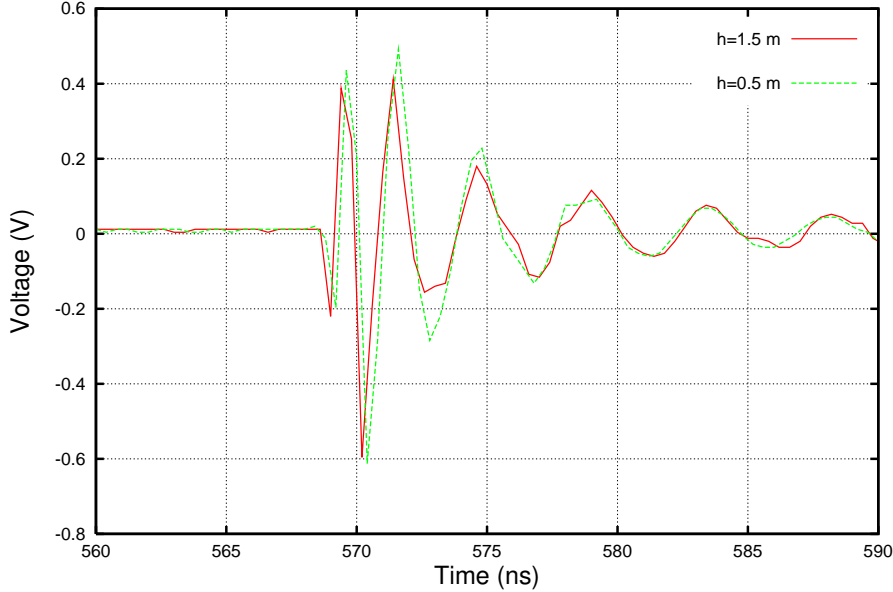}} \caption{Comparison of broadcasting in-air (VV) with h=0.5 meters vs. h=1.5 m, corresponding to antennas resting on the snow surface, and elevated by one meter, respectively. The absence of large differences between the two signals indicate that interference effects between the direct path and the path reflected off the air-snow interface are not substantial. Note that, for h=1.5 m, the surface
reflection is approximately one beamwidth off boresight.} 
\label{fig:TD-h=1.5m-vs-.5m-S12air.eps} 
\end{figure}

In principle, the characteristics of the 
received through-air signal are polarization dependent.
There may be interference between a direct-path and an 
optical path corresponding to reflection off the snow surface.
Defining ``V'' here as the vertical axis perpendicular to the surface 
(${\hat z}$), and ``H'' as the horizontal
axis perpendicular to both ``V'' and 
the propagation vector, 
we note that the magnitude of interference in VV vs. HH will differ, since
the electric field vector components parallel and perpendicular to the
air-ice interface have different, angle-dependent reflection
coefficients, as prescribed by the Fresnel equations.
At the Brewster angle (tan $\theta_B=n_{ice}/n_{air}$), e.g., the
transverse amplitude approaches zero
and the reflected angle is longitudinally polarized.
Such effects can, in principle, be probed by varying the separation 
distance between transmitter and receiver over a scale comparable 
to a wavelength. To assess interference,
we made a set of measurements with the 
two antennas relatively close to each other (d=5 m), and
both with and without 1-meter high elevating ice blocks.
In general, the time difference between the direct path and the air-surface
reflected path is given by: $c\Delta t=2(\sqrt{h^2+d^2/4}-d/2)\approx
2h^2/d$. 
Assuming the phase center of the antenna is at a height of 0.5 meters,
corresponding to no ice blocks, we obtain
$\Delta t\sim0.333$ ns. For the case where the 
antennas are elevated by an
additional 1 
meter, then $\Delta t\sim$3 ns, corresponding to approximately a
phase difference of 3$\pi$ radians at 500 MHz.
%@300 MHz, lambda=1m, T=lambda/c=3.333 ns. At 500 MHz, T=2 ns
Figure \ref{fig:TD-h=1.5m-vs-.5m-S12air.eps} shows the 
comparison for these two elevations. The magnitude of the observed
differences indicates that interference between the two possible
paths is not substantial. This conclusion is also 
supported by Figure \ref{fig:TD0.eps}, which shows good 
agreement between the signals observed for the two polarizations
(albeit with a phase shift of $\pi$ radians) in the elevated configuration.

\subsubsection*{Comment on surface propagation}
It has been suggested\citep{r:ralston05} that ``surface
waves'', or evanescent solutions to Maxwell's Equations, may offer
a promising technique for detection of neutrino-induced
radiowave signals. In this model, neutrino-ice collisions in the
ice sheet result in radio waves directed towards the surface,
which then propagate along the surface with relatively little
attenuation, towards suitably located surface receivers.
Such z-polarized (Vpol, in our
case) signals would be
`trapped' on the air-ice boundary and therefore have
an amplitude dependence varying as $\sqrt{r}$. Our ${\rm S}_{12}$
measurements indicate no preference for Vpol vs. Hpol, and
are more consistent with spherical 1/r amplitude diminution.

\subsection*{${\rm S}_{12}$(ice) measurements - signal amplitude}
Antennas were initially oriented relative to the local 
elevation gradient, which
is presumed to coincide with the local ice flow axis, 
pointing in the direction
of Taylor Valley, approximately 50 km away.
%http://www.cs.uaf.edu/~bueler/icemath.pdf

\subsubsection*{Estimate of ice depth at Measurement Site}
\begin{figure} \centerline{\includegraphics[width=12cm]{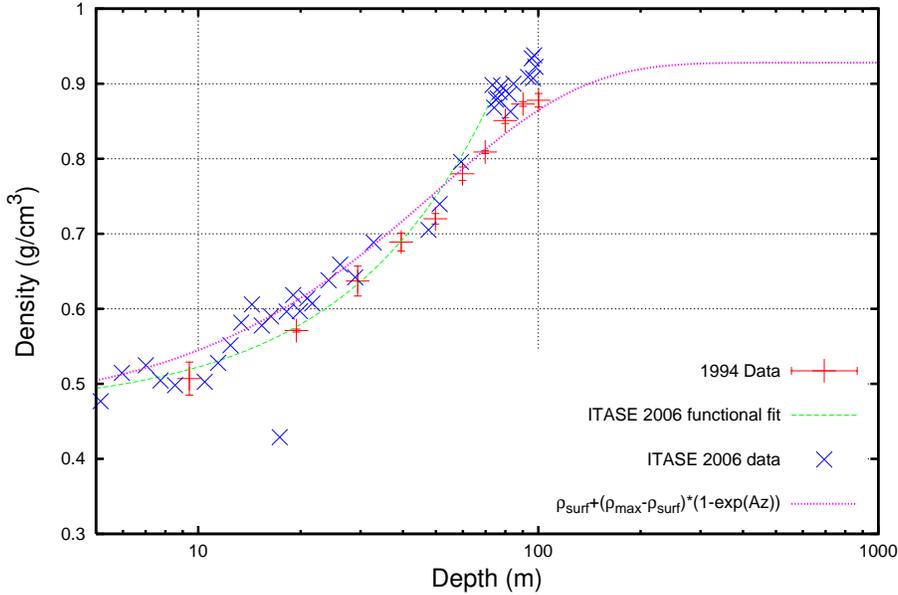}}%{TD-rhovsz-tobed.eps}} 
\caption{Density data, taken 
at Taylor Dome\citep{Fitzpatrick-1994},
at the our borehole site\citep{Dixon-2007},
and functional form of assumed
density profile, extrapolated to bedrock.}\label{fig:TD-rhovsz-tobed.eps} 
\end{figure}

To determine the attenuation length, we must first estimate the ice depth.
From the time delay between the sent and the received signal,
we can estimate the
depth of the ice shelf at the measurement site. We use 
Taylor Dome ice density data tabulated elsewhere\citep{Fitzpatrick-1994}
%http://earthweb.ess.washington.edu/\~bo/papers/01-109/01-109.pdf
to account for the variation in wavespeed with density,
and assume an exponential
density profile beyond 100 m, as shown in Figure
\ref{fig:TD-rhovsz-tobed.eps}.
Also overlaid are preliminary data
provided by the ITASE
group based on cores 
extracted at the time of the November, 2006 drilling\citep{Dixon-2007}. 
Given the errors, we take these data as qualitatively consistent
with the more precise Taylor Dome density data.
Comparing the travel time difference for the in-air 
vs. in-ice measurements (cable delays are the same for both), and
knowing the in-air tabulated separation distances, the cable delay
offset can be cancelled. Calculating an in-ice transit time of
11100 ns (corrected for cable delays, and based on the
average of the HH and VV signal return times; the discrepancy
between these corresponds to a depth uncertainty of order
2 meters), %dt(ice)-dt(air) - (d/c,air) 
we obtain an 
estimated depth of 940$\pm$15 m, where the systematic error 
shown is dominated by the unknown density profile below 100 m.
This value is consistent with other estimates, albeit at 
$\sim$5 MHz\citep{Welch-2006}. (In principle, the equality of the signal
propagation times at 5 MHz vs. 500 MHz can be used to bracket the
dispersive characteristics of ice over two decades in frequency. The 
timing resolution $\delta t$, however, is expected to follow
$\delta t\sim 1/f$.)
Knowing the pathlength of the radiofrequency signal through ice,
we can now use this value to extract the 
average ice attenuation length $L_{atten}$.

%Data: for 5 m S12(air), t0=105 ns, t(Rx)=567 ns, transit time=5m/c=16.66 ns
%for S12(ice), t0=122 ns, t(Rx)=11715 ns, transit time=(11715-122)-(567-16.66-105)=11148 ns. 
%d=ct=>d=11.148e-6*2.991e8/(2.*1.7) - 

\subsection*{Numerical Extraction of Attenuation Length}
\subsubsection*{Comparison of ${\rm S}_{12}$(ice) to ${\rm S}_{12}$(air) power spectra}
\begin{figure} \centerline{\includegraphics[width=12cm]{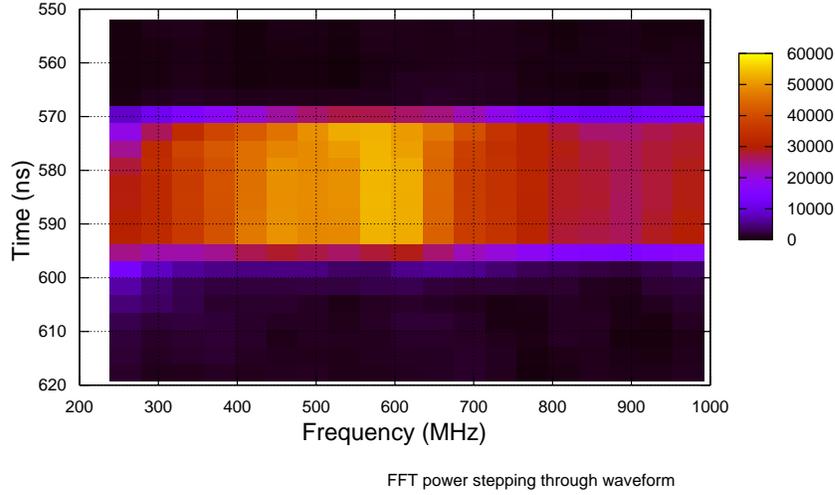}}%{ft-absol-pm3d-zoom-sig0.eps}} 
\caption{V(f) characteristics of ${\rm S}_{12}$(air) signal
(400 ps/sample). z-axis has dimensions [V/MHz]; absolute normalization is arbitrary. To obtain this plot, we have calculated the Fourier
transform of V(t) for the through-air received pulser output signal as we step through the waveform.} \label{fig:ft-absol-pm3d-zoom-sig0.eps} 
\end{figure}

\begin{figure} \centerline{\includegraphics[width=12cm]{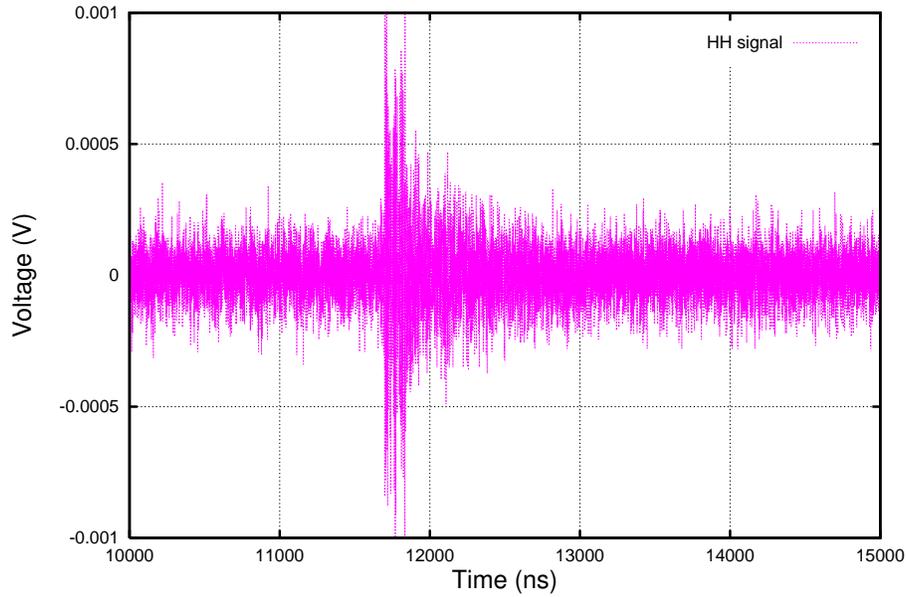}} \caption{Typical observed HH reflection signal.} \label{fig:HH-1165563686-full.eps} 
\end{figure}

\begin{figure} \centerline{\includegraphics[width=12cm]{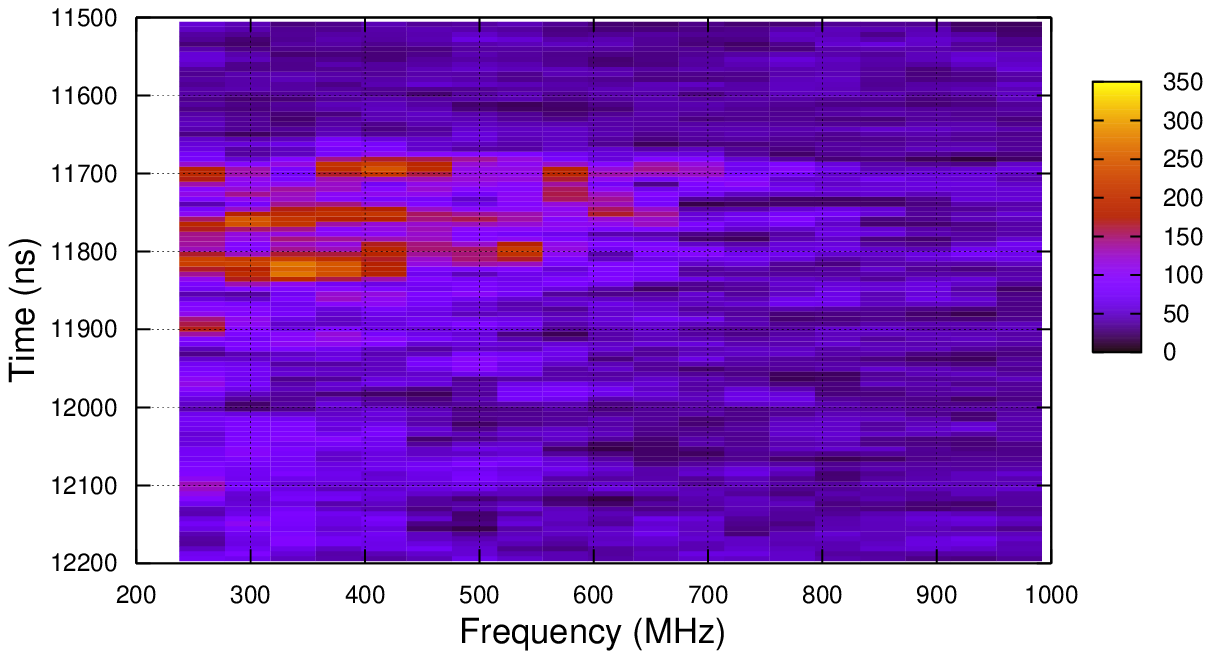}} \caption{VV Reflected signal spectral power.} \label{fig:ft-absol-1165574302_VV.eps} 
\end{figure}

\begin{figure} \centerline{\includegraphics[width=12cm]{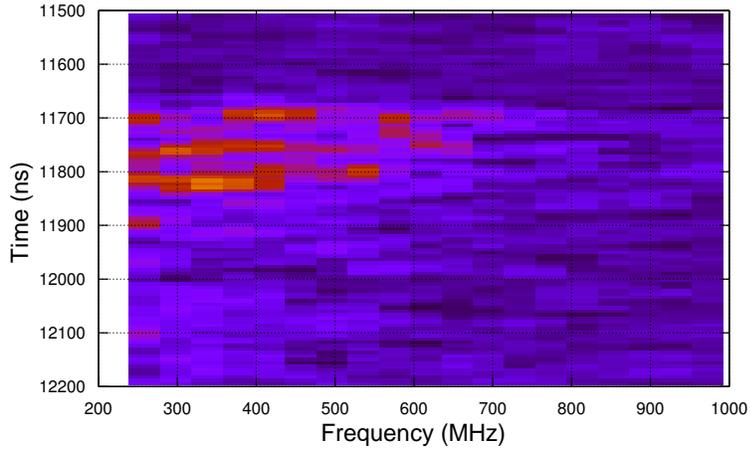}} \caption{HH Reflected signal spectral power.} \label{fig:ft-absol-1165563686_HH.eps} 
\end{figure}

\begin{figure} \centerline{\includegraphics[width=12cm]{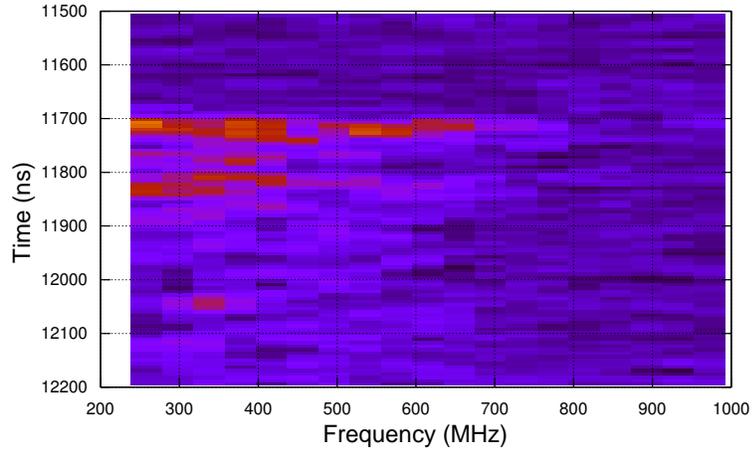}} \caption{'Rotated' VV reflected signal spectral power. Here, the antenna is rotated in the horizontal plane from the initial VV orientation into the HH orientation.} \label{fig:ft-absol-1165565470_VV.eps} 
\end{figure}

\begin{figure} \centerline{\includegraphics[width=12cm]{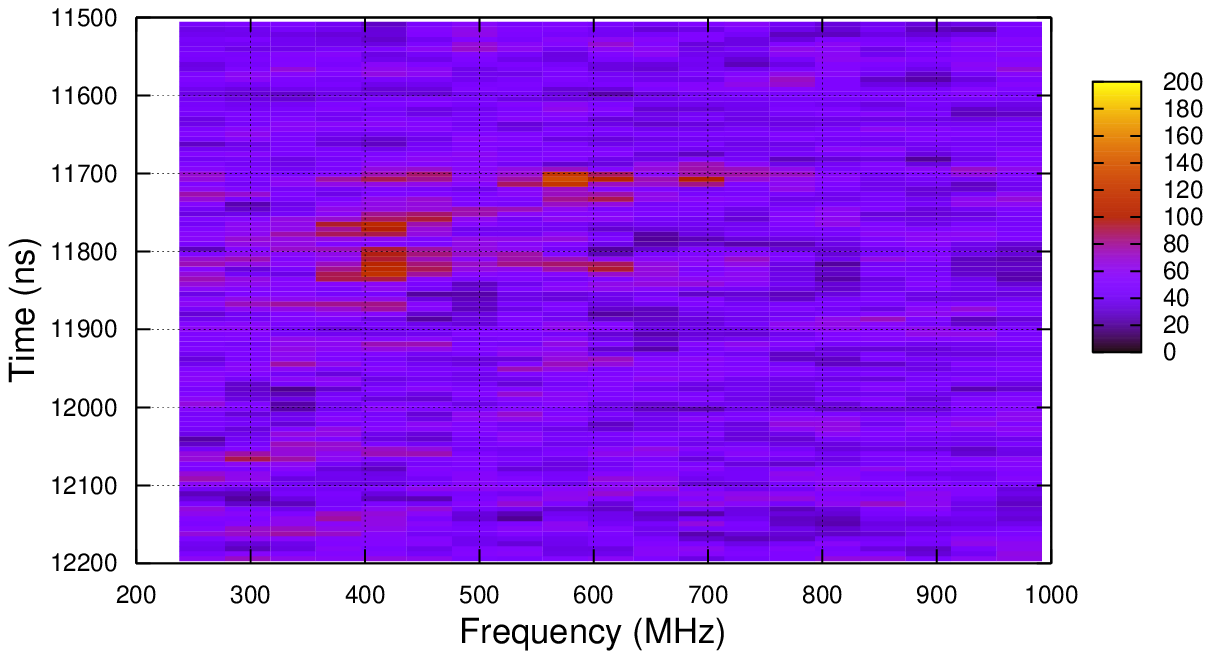}} \caption{VH 'cross-polarization' reflected signal power.} \label{fig:ft-absol-1165567212_VH.eps} 
\end{figure}

We compare the signal observed over the 
path-length of the in-ice signal
with the signal observed through-air. 
Figure \ref{fig:ft-absol-pm3d-zoom-sig0.eps} shows the 
power spectrum of the through-air signal.
We observe considerable power at high frequencies (up to 1 GHz), over
a timespan of 10 ns.
By contrast, the through-ice received signal
is apparently extended by nearly two orders of magnitude. 
Figure \ref{fig:HH-1165563686-full.eps} shows a typical
reflection signal.
%\begin{figure} \centerline{\includegraphics[width=12cm]{ft-absol-pm3d-zoom.eps}} \caption{V(f) of signal region} \label{fig:ft-absol-pm3d-zoom.eps} \end{figure}
Additional
V(f) plots for the through-ice signals are shown in
Figures \ref{fig:ft-absol-1165574302_VV.eps}--\ref{fig:ft-absol-1165567212_VH.eps}.
We typically observe similar spectral content, compared to the ${\rm S}_{12}$(air) signal,
only for the first 50 ns of received signal.
Our power spectra data are also qualititatively consistent with 
both ice absorption 
increasing with frequency, as well as the 
$1/\lambda^4$ dependence characteristic of Rayleigh scattering.

%for the initial VV configuration, Figure \ref{fig:ft-absol-pm3d-zoom-sig0.eps} for the initial HH configuration, Figure \ref{fig:ft-absol-1165565470_VV.eps} for the 'rotated' VV configuration, and Figure \ref{fig:ft-absol-1165567212_VH.eps} for the VH 'cross-polarization' configuration.

%http://www.stolaf.edu/other/cegsic/publications/Miners%20et%20al.PDF

Uncertainties
in the frequency dependence of electromagnetic wavespeed in polar
ice are also sufficient to allow dispersive effects as a possible
explanation for the dilated received signal structure. In that case,
our data would be consistent with $dn/d\omega<$0 in the frequency
range of interest.
However, it should be noted that, within the limits of
our timing resolution ($\sim$100 ns),
our previous experiment at South Pole\citep{Bar2005}
showed very similar signal arrival times for 320 MHz vs. 700 MHz
over a total propagation time of 33$\mu$s.
By contrast, our Taylor Dome
data indicate received
signal durations of order 120 ns, relative to a total
transit time of 11.1~$\mu$s. 

We first extract the attenuation length using the relative signal
strengths observed for in-air vs. in-ice broadcasts.
To minimize relative 
distance measurement error, we 
normalize the ${\rm S}_{12}$(ice) signal
voltage to the ${\rm S}_{12}$(air) signal measured for maximum in-air separation.
We compare the calculated attenuation lengths
based on: a) the first 10 ns of
enhanced voltage amplitude observed in the reflected signal $V_{max}$,
b) the calculated integral of the signal strength, using 50 ns of
both in-ice and in-air data, based on the similarity of the
broadcast vs. received
frequency spectra over this timescale, and c) integrating 250 ns after the
initial observed reflection appears to include the entire observed
reflection enhancement.
For these estimates, we must
directly subtract the average voltage $\langle V\rangle_{pre-signal}$, as
measured for one microsecond just before the arrival of the reflected
signal. Additionally, the ambient rms voltage $\sigma_{V,pre-signal}$ must
be subtracted in quadrature. 
The corrected calculated signal voltage
$V_{signal}=\sqrt{(V_{measured,scope}-\langle V\rangle_{pre-signal})^2-\sigma_{V,pre-signal}^2}$.

Alternatively, these values are determined from direct
application of the Friis Equation, as outlined previously.
%Figure \ref{fig:TD_calc_Latten_final.eps} shows the distribution of extracted attenuation lengths, obtained from several (${\rm S}_{12}$(air),${\rm S}_{12}$(ice)) combinations. Integrating over 50 ns of signal yields an average attenuation length of 460 m; the rms of those values is approximately 27 m. Using the peak voltage only, we obtain $L_{atten}=443\pm25$ m. \begin{figure} \centerline{\includegraphics[width=12cm]{TD_Latten.eps}}%TD_calc_Latten_final.eps}} \caption{Ensemble of estimated values of field attenuation length, normalizing HH and VV ${\rm S}_{12}$(ice) measurements to ${\rm S}_{12}$(air) measurements at separation distances of 50 m and 60 m.} \label{fig:TD_calc_Latten_final.eps} \end{figure}
%This is considerably less attenuation compared to a recent measurement made on the West Antarctic Ice Sheet (http://www.ig.utexas.edu/people/staff/gcatania/winebrenner_catania.pdf?PHPSESSID=def1b9), which yielded field attenuation lengths in the range 240-320 m.

\subsubsection*{Uncertainties in surface scattering at the bedrock}
The nature of the bottom reflection critically affects the magnitude of the
received signal.
In the extreme case, one might assume a mirror-like bottom reflectivity
of  unity. A more realistic
value of reflection
coefficient, corresponding to a typical rock-ice dielectric contrast, is
approximately 0.3. 

\subsubsection*{Average bottom slope and
BEDMAP elevation data.}
A bottom surface which is linear in elevation slope, either positive or
negative, has the effect of reducing the net estimated signal, as does a
concave upwards surface (i.e., a local 'hill'). 
A concave downwards bedrock surface (i.e., a
local `depression') could have a focusing effect. In our
calculation, we have ignored such possible
curvature effects.
The BEDMAP group has tabulated estimated bedrock\citep{BEDMAP} and ice
surface elevations across the Antarctic continent. Over
5 km$\times$5 km grid squares in the vicinity of our measurement
site, surface and bedrock elevations are shown in 
Table \ref{t:elev}. 
The average slope of the
bedrock over the area where
our measurements have been made is very slight, of order 5 mrad,
%30 m/5000 m
although local variations exceeding that cannot be ruled out
without further measurements. \message{WHAT ARE BEDMAP ELEVATION ERRORS?}
\begin{table}[htpb]
\caption{Surface/Bedrock elevations (BEDMAP). Borehole site has
coordinates $\approx$ (1 km,2 km) relative to given grid center
coordinates. \label{t:elev}}  %Elevation measurement errors are  10 m.
\begin{tabular}{c|c|c}
x0 (km) & y0 ($\equiv$Grid North) (km) & Surface/Bedrock elevation (m) \\ \hline
-5 & -5 & 2232/1491 \\
-5 & 0 & 2185/1372 \\
-5 & 5 & 2194/1302 \\
0 & -5 & 2305/1512 \\
0 & 0 & 2275/1424 \\
0 & 5 & 2291/1379 \\
5 & -5 & 2350/1464 \\
5 & 0 & 2353/1422 \\
5 & 5 & 2353/1393 \\ \hline
\end{tabular}
\end{table}

\subsubsection*{Bedrock features}
The signal magnitude
is sensitive to the scale of the roughness (i.e., surface features, or
inhomogeneities) at the bedrock.
We expect that, if the bottom surface were characterized by
roughness of typical vertical scale $h$, then 
the observed signal should be specular and completely
coherent for $\lambda\gg h$ and show a
loss of coherence for $\lambda\ll h$. 
In the former case,
the specular signal from the horn transmitter
is coherent over
a bedrock
disk (``Fresnel zone'') 
of radius $R\sim\sqrt{2\lambda d_{Tx}d_{Rx}/(d_{Tx}+d_{Rx})}/2$,
corresponding to the region over which the phase
variation is $<\pi/2$ radians. %sqrt(d1*d1+r*r)+sqrt(d2*d2+r*r)-d1-d2<lambda/2 
Here,
$d_{Tx}=d_{Rx}$ is the distance from the transmitter or
receiver through the ice to the bedrock and $\lambda$ is the 
observed wavelength. For $d\sim$1000 m and $\lambda\sim$1 m,
then $R\sim$30 m, $R\sim\sqrt{\lambda d_{Tx}}$,
corresponding to a maximum smearing in the time domain of approximately
2 ns due to the non-zero size of the illuminated disk.
Beyond that region, there will be coherent Fresnel zones at larger
radii, with
an expected diminution in amplitude.

\begin{figure} \centerline{\includegraphics[width=12cm]{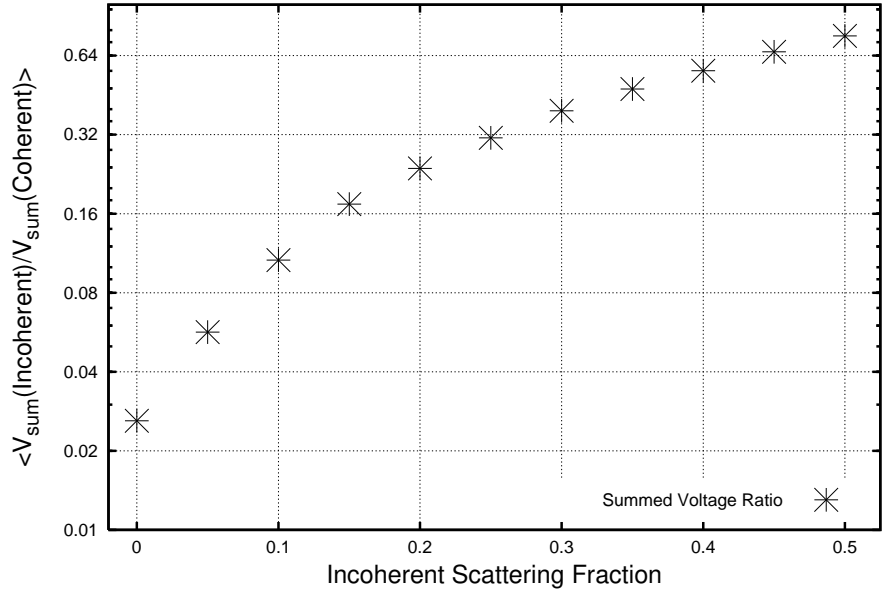}} \caption{Expected ratio of integrated signal amplitudes assuming coherent vs. incoherent scattering at bedrock. Fraction of power scattered incoherently is plotted horizontally.}\label{fig:Coherent_model.eps} 
\end{figure}

In the
idealized limit of completely incoherent
scattering (we assume no depolarization), the
high-frequency, low-wavelength power would scatter with a $1/R^4$
dependence off the bottom, vs. a
sharper $1/R^2$ dependence for 
coherent specular reflection. 
The observed time duration of the
reflected signal (Figure
\ref{fig:HH-1165563686-full.eps})
implies an illuminated region roughly $10\times$ larger
than our 30-meter estimate above.
For incoherent scattering, we expect signal to arrive at the
receiver over a timescale dictated by the beamwidth in-ice
($\pm$12 degrees), corresponding to approximately 250 ns given the
measured bedrock depth. 
Our data therefore suggest a substantial
fraction of 
incoherent scattering
at the bedrock.
To assess the signal
amplitude reduction due to loss
of coherence, we have calculated the voltage
reduction as a function of the scale of the surface scattering 
at the bedrock,
as illustrated in Figure
\ref{fig:Coherent_model.eps}. This scale, relative to the typical
wavelength scale, determines the fraction of signal
which scatters coherently vs. incoherently at the bedrock.
For the calculations of attenuation length
presented below, our ``Incoherent''
assumption corresponds to an incoherent fraction of
0.25.

\subsubsection*{Previous Bottom Reflection Data}
\begin{figure}\centerline{\includegraphics[width=12cm]{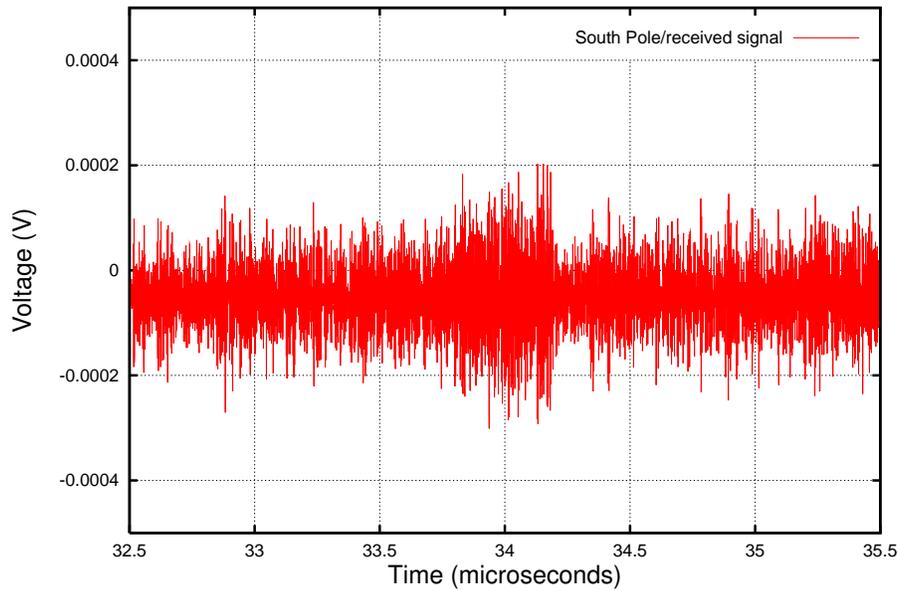}}\caption{Time domain (V(t)) signal observed in bottom reflection measurement at South Pole, January 2004.} \label{fig:SP-380MHz-400ns.eps} 
\end{figure}

\begin{figure} \centerline{\includegraphics[width=12cm]{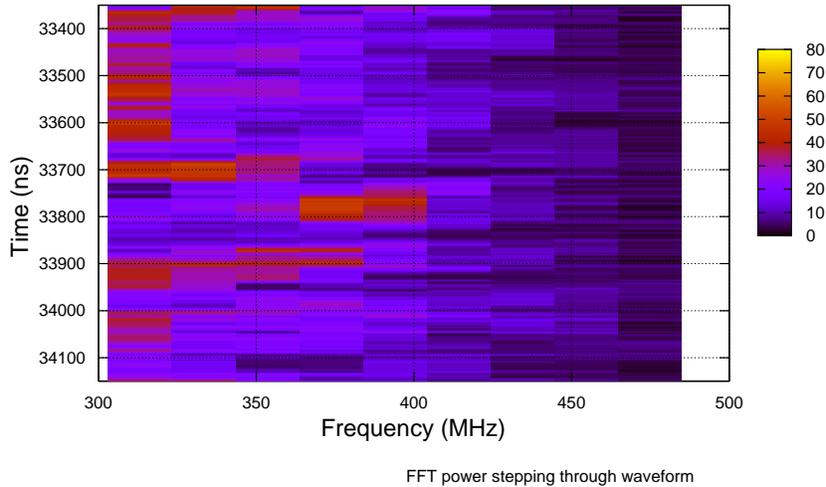}}\caption{380 MHz Signal observed in bottom reflection measurement at South Pole, January 2004. Signal duration 40 ns.} \label{fig:380MHz-40ns-pulse-spole.eps} 
\end{figure}

We have also used previous data, taken at the South Pole in January, 2004
with similar apparatus, for possible evidence of an extended 
reflected signal time duration relative to the transmitted signal\cite{Bar2005}. 
(Since the antenna orientation relative to ice flow was not noted,
and since data were only taken in the ``co-pol'' alignment, with
Tx and Rx signal polarization axes aligned parallel with each other,
or ``cross-pol'', for which the receiver axis (only) was rotated by 90 
degrees on the surface, these data cannot be used to extract
birefringence results, discussed later in this document.)
The reflection from a 
monochromatic 320 MHz, 400 ns long `tone' signal emitted from
a TEM horn transmitter, then later registered in a surface
horn receiver (Figure \ref{fig:SP-380MHz-400ns.eps}) is observed to be
approximately 400 ns in duration. A narrower input transmitter signal (40 ns)
at a slightly higher
frequency (380 MHz) similarly results in a correspondingly narrow received signal (Figure \ref{fig:380MHz-40ns-pulse-spole.eps}).
Although not conclusive,
this observation is consistent with our interpretation that the primary
received reflection is of comparable duration, and has comparable spectral
content, as the broadcast signal. This indicates a coherent, specular
reflection off the underlying bedrock at South Pole.
%The power spectra data (Figure \ref{fig:jan27_2004_SP_copol_320MHz.eps} and Figure \ref{fig:jan27_2004_SP_xpol_320MHz.eps}), unfortunately, do not allow us to conclusively assess a possible time delay between the two measured polarizations.

%\begin{figure} \centerline{\includegraphics[width=12cm]{jan21_nobalun_copol_380MHz.eps}}\caption{Raw time domain signal (V(t)) observed in bottom reflection measurement at South Pole, January 2004.} \label{fig:SP-380MHz-400ns.eps} \end{figure}

%\begin{figure} \centerline{\includegraphics[width=12cm]{jan27_2004_SP_copol_320MHz.eps}} \caption{``Co-polarization'' signal observed in bottom reflection measurement at South Pole, January 2004. Absolute magnitude is arbitrary, with dimensions [V/MHz].} \label{fig:jan27_2004_SP_copol_320MHz.eps} \end{figure}

%\begin{figure}\centerline{\includegraphics[width=12cm]{jan27_2004_SP_xpol_320MHz.eps}}\caption{``Cross-polarization'' signal observed in bottom reflection measurement at South Pole, January 2004. Here, the receiver has been rotated by 90 degrees relative to the transmitter alignment in the previous Figure.} \label{fig:jan27_2004_SP_xpol_320MHz.eps} \end{figure}

\subsubsection*{Internal Layering Effects and Volume Scattering}
%http://www.geophys.washington.edu/Surface/Glaciology/PROJECTS/SIPLE/sdmall.gif
%http://www.nsf.gov/pubs/1999/nsf98106/98106htm/ht008.html
%http://www.agu.org/pubs/sample_articles/cr/2002GL016403/2002GL016403.pdf
%websearch on ``Antarctica radargram''
%http://www.urova.fi/home/hkunta/jmoore/gpr_cryo.pdf
%http://tornado.rsl.ku.edu/2002pdf.htm
% kenny - Fabric starts to develop several hundred meters.  With a large scale, it is gradually changed.  However, we can see rapid alternations.  See Obbard and Bkaer in J Glaciology vol. 53 p.41- (2007).
Layers can arise as a result of episodic
events (dust or acid layers, e.g., due to 
volcanic eruptions\citep{Millar-1981}, which result in a change in
ice conductivity (and therefore $\epsilon''$) or annual processes,
including thin surface crusts which form in the summer,
constituting a discontinuity in density
and resulting in
a contrast in the real part of the dielectric constant\citep{Bogorodsky,Smith&Evans,Fujita-1999}. 
Layers of the
latter type should become less important with
depth as the ice approaches its asymptotic density.
Anisotropies in the crystal structure of the ice may also
constitute a 
discontinuity. Nearer 
to the coast, brine infiltration can also result in
stratified layering. 
%However, this is expected to be negligible at the South Pole.
In magnitude,
the ground-penetrating
radar (GPR)
returns from internal layers are typically 50-100 dB smaller
than the returns from the bottom ``echo''.
Precise study of layers requires extremely sensitive receivers and, in
the case of annual layering,
the ability to distinguish density differences of order
1-10 cm apart ($<$1 ns) in the ice.
Interferometry or pulse modulation
techniques can also be used to probe layering, provided the 
interferometer is sensitive to 1 ns time scales.
In addition to discrete layers, quasi-continuous $\lambda^{-4}$
Rayleigh scattering off air bubbles in unenclathrated ice has been
considered quantitatively\citep{Smith&Evans}, resulting in
a `worst-case' attenuation loss estimate of
0.7$\lambda^{-4}$ dB per 100 m. %Bogorodsky, p. 176
This extreme case corresponds to an amplitude loss of a factor of $\sim$6,
for $\lambda_{ice}\sim$1 m ($f_{air}\sim$500 MHz).

\subsubsection*{Scattering Effects in Data}
\begin{figure}
\centerline{\includegraphics[width=12cm]{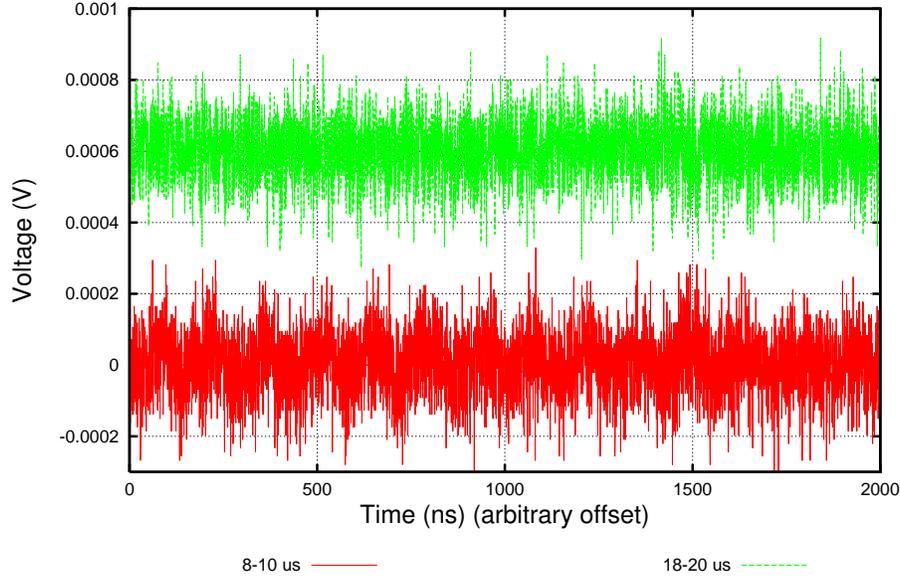}}\caption{Comparison of pre-reflection waveform with post-reflection
waveform, no frequency filtering. Traces have been vertically offset for visual clarity.}
\label{fig:pre-post.eps}
\end{figure}

%Following added after paper submitted to J. Glac.:
%There have been, nevertheless, extensive measurements of the conductivity of ice cores in the DC limit, as well as the $10^5$ Hz regime\footnote{See, e.g., http://nsidc.org/data/waiscores/pi/taylor.html.}, as indicated in Figure \ref{fig:DEP-Eisen-B2-2000.eps}, which provide additional information on internal layering.
Pre-signal reflections in our captured time-domain waveforms which
should result from either internal layering or volume scattering are
not immediately evident in our data.
Comparison of the waveform in the interval
8-10 $\mu$sec from the trigger time with the 
interval 18-20 $\mu$sec from the trigger time
shows that the rms voltage is 
actually slightly smaller in the former %040728
time interval than the latter, inconsistent with quasi-continuous
Rayleigh scattering (Figure \ref{fig:pre-post.eps}).
The rms values in these plots ($\sim$100 $\mu$V) 
are roughly consistent with the expected thermal noise rms voltage $V_{rms}\sim\sqrt{kTBZ}\sim$20 $\mu$V, and the amplification (20 dB in power) and coaxial cable losses between the receiver and the data acquisition system.

\begin{figure}[h]
\begin{minipage}{18pc}
\vspace{1cm}
\includegraphics[width=7.5cm]{fft-1165574302_VV.eps}
\vspace{0.2cm}
\caption{V(f) of reflection signal waveform, over region prior
to apparent signal reflection.}
\label{fig:fft-1165574302_VV.eps} 
\end{minipage}
\hspace{0pc}
\begin{minipage}{18pc}
\includegraphics[width=8cm]{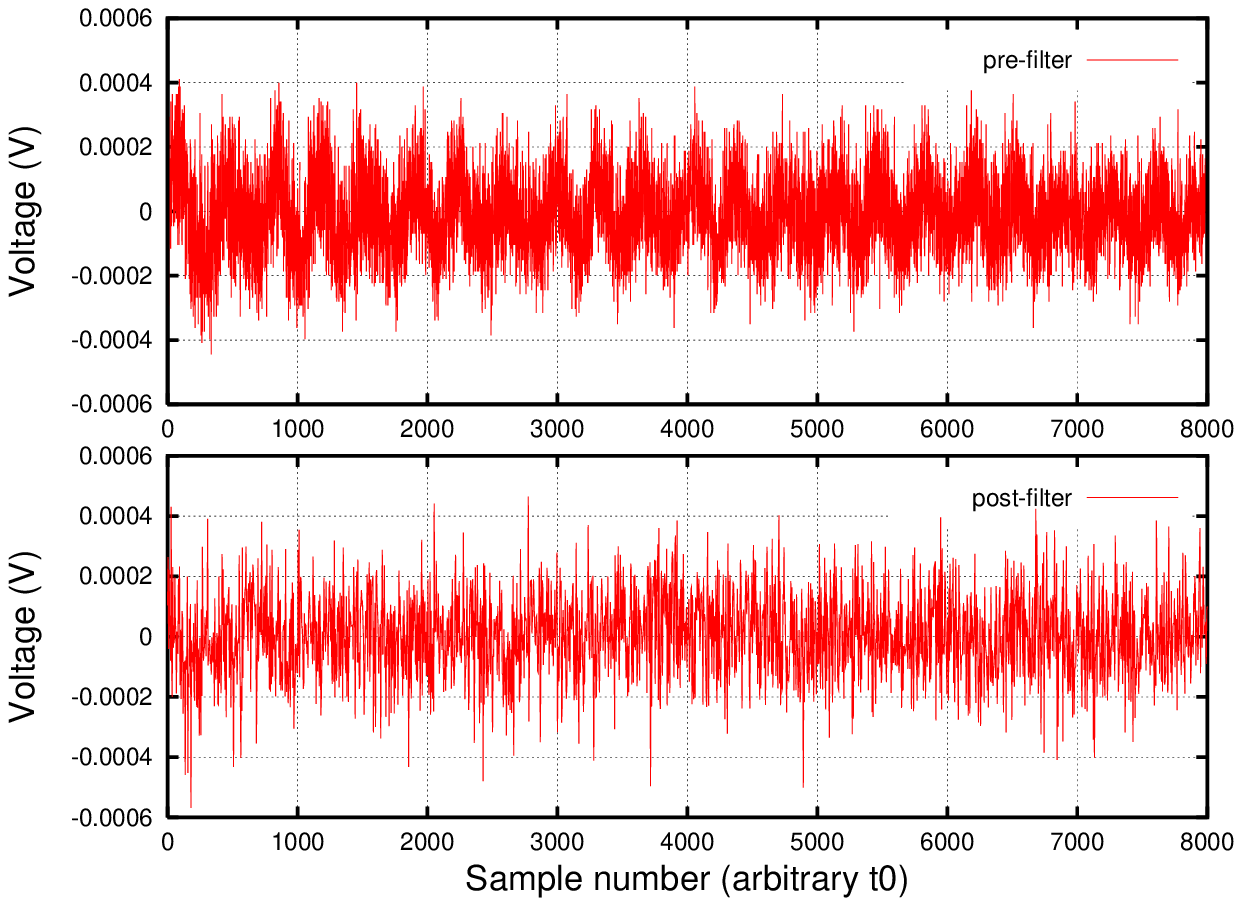}
\caption{Comparison of pre- vs. post-filter waveform, in 
region prior to apparent signal reflection.}
\label{fig:pre-post-filter-1165574302.eps}
\end{minipage} 
\end{figure}

We observe a clear modulation of the pre-reflection signal, on a
timescale of several hundred ns. %$\sim$150 ns. 
After digitally filtering out this
modulation from the Fourier transform (Figure \ref{fig:fft-1165574302_VV.eps}),
the remaining signal appears more `thermal' in its appearance
(Figure \ref{fig:pre-post-filter-1165574302.eps}).
The source of this apparent modulation of the pre-reflection signal
is not understood, however, we note that it does not contribute
substantial power to the waveform and should therefore not substantially
affect our attenuation length estimates.

\begin{figure}\centerline{\includegraphics[width=12cm]{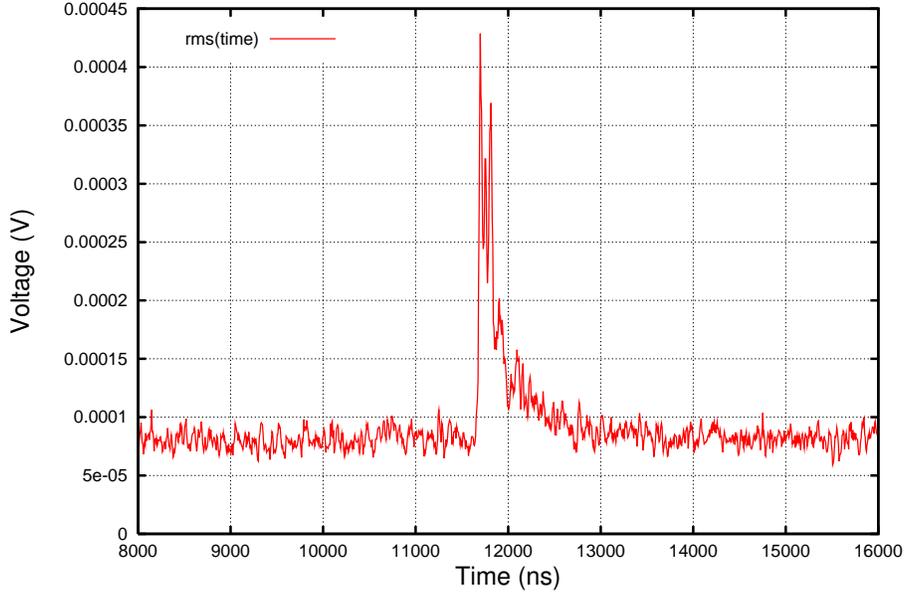}}\caption{rms Voltage as a function of time, original VV Tx and Rx configuration. Power below 32 MHz has been digitally removed from original V(t).}\label{fig:TD-rms-V-time.eps} 
\end{figure}

\begin{figure} \centerline{\includegraphics[width=12cm]{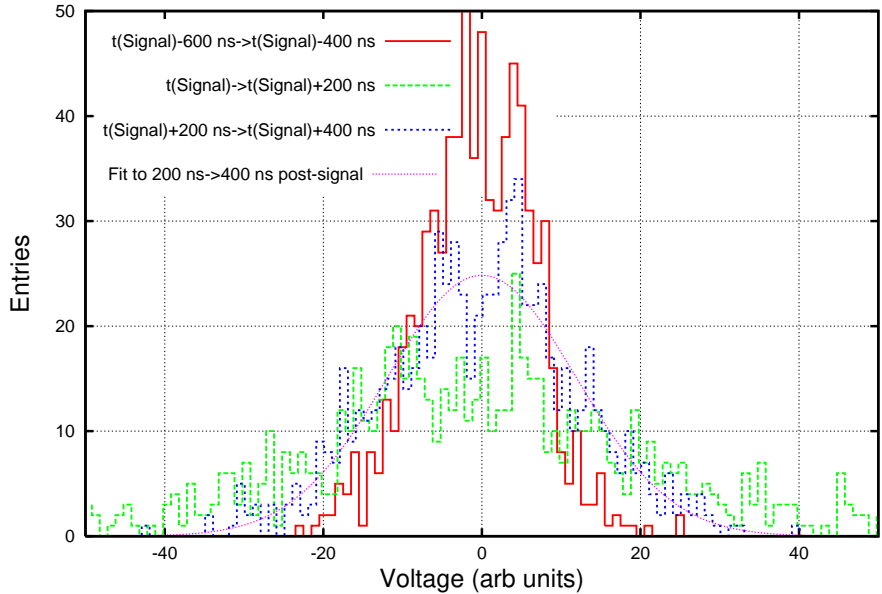}} \caption{Voltage distribution around the reflection signal and fit to Gaussian distribution. Presence of non-Gaussian `tails' indicate non-thermal effects.} \label{fig:Vhistfile_1165563686.eps} 
\end{figure}

Figure \ref{fig:TD-rms-V-time.eps} displays the rms voltage as a function of time for the original VV configuration. We note again that the regions prior to and after the signal reflection are generally consistent with each other. 
All power below 32 MHz has been filtered out of this plot prior to calculation of the rms. %1165574302_VV.gnudat
Dividing the waveform
shown in Figure \ref{fig:HH-1165563686-full.eps} into 200 ns long segments
beginning i) 600 ns after the apparent signal reflection, ii) at the
time of the apparent signal reflection, and iii) 200 ns after the
apparent signal reflection, we observe that the distribution of
voltages in the waveform show excursions beyond those expected
for a simple
Gaussian distribution, as would otherwise be expected for pure thermal noise 
(Figure \ref{fig:Vhistfile_1165563686.eps}).

%Qualitatively, the signals displayed the same time-domain structure, and the same peak voltages (Figure \ref{fig:draw_pol_phi_scan.eps}) at the second (displaced) location as the initial location. 

%Correcting for the 20 dB amplifier gain, we have, for a 940 meter depth: 0.14 mV = $(24000/1910)\cdot exp(-1910/L_{atten})$, or $L_{atten}\sim$422 m, or $\sim$10.3 dB/km in amplitude. 

Assumption-dependent numerical results for our 
pathlength-averaged
attenuation length are presented in Table \ref{tab:Lattensum}. 
Results are presented for a variety of assumptions
regarding the coherence of the signal at the bedrock, as well as the
reflection coefficient at the bedrock interface. 
\begin{table}[htpb]
\caption{Summary of attenuation length measurements, under various
assumptions for the reflection coefficient and the coherence characteristics
of the underlying bedrock. Values shown are averaged over multiple
measurements. Errors represent statistical spread of calculated values
only. \label{tab:Lattensum}}
\begin{tabular}{c|c|c|c|c} 
Assumed & Signal & Assumed & Integrated & Calculated \\
Reflection Coeff. & Normalization & Basal Scattering & Reflected Signal & 
$\langle L_{atten}\rangle$ \\ \hline
1.0 & In-air & Coherent & 10 ns & $340\pm15$ m \\
1.0 & In-air & Coherent & 50 ns & $351\pm15$ m\\
1.0 & In-air & Coherent & 250 ns & $616\pm32$ m\\
1.0 & In-air & Incoherent & 10 ns & $441\pm25$ m\\
1.0 & In-air & Incoherent & 50 ns & $458\pm26$ m \\
1.0 & In-air & Incoherent & 250 ns & $1055\pm95$ m \\ \hline
1.0 & Absolute & Incoherent & 250 ns & 628 m \\
0.3 & Absolute & Incoherent & 250 ns & 1051 m \\ \hline
\end{tabular}
\end{table}

In principle, one might argue that one should properly include the
cross-polarization (i.e., HV or VH) signals into the tabulation of the
received signal. Reciprocity suggests that, if a purely V-transmitted
signal can excite the H-terminals of the receiver
horn antenna, excitation of the V-terminals of the transmitter can also
result
in excitation of the H-terminals of the transmitter. In that
case some of the H-received signal is simply the transmission of this
H-component, which has preserved its polarization in its traversal of the
intervening ice. We therefore only include signal from that polarization 
component corresponding to that originally transmitted. Including the
cross-polarization signal would increase the estimated attenuation length
by $\sim$10\%. 
%(Unfortunately, a complete set of VV, HH, VH, and HV measurements were not taken.)

\subsection*{Attenuation length dependence on temperature profile}
\begin{figure} \centerline{\includegraphics[width=12cm]{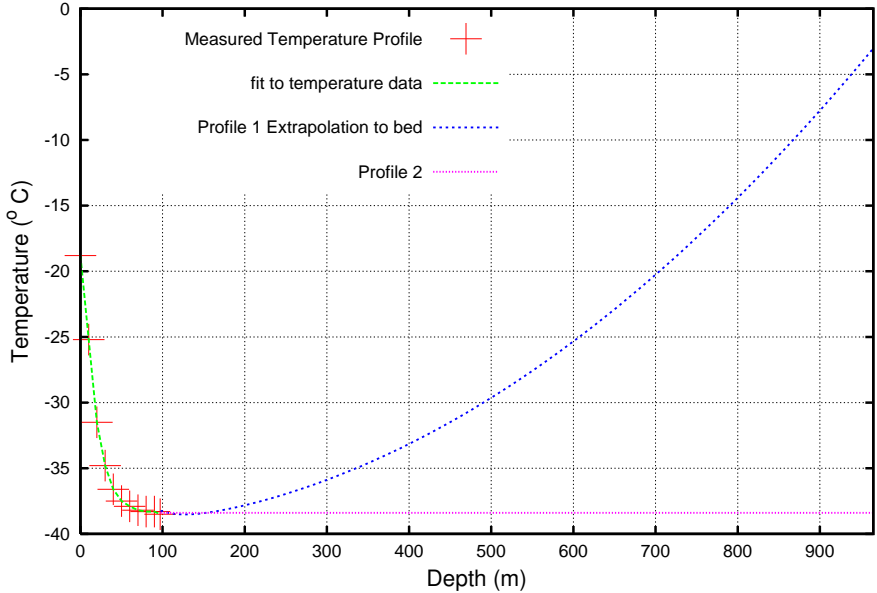}} \caption{Measured temperature profile at borehole site.
Also shown are 
temperature profiles considered for systematic uncertainty estimate in
extracted attenuation length. We note that all measurements (to our
knowledge) indicate a warming of ice to near 0$^\circ$ C at the bedrock;
``Profile 2'' is therefore considered to be an extremely
unlikely, and overly pessimistic case.} \label{fig:TD-tvsz.eps} 
\end{figure}

Radiofrequency absorption is a strong function of temperature.
Figure \ref{fig:TD-tvsz.eps} displays our measurements of temperature
as a function of depth at our borehole site, overlaid with two possible
extrapolations in the region below our last temperature measurement
at a depth of 100 meters. 
We estimate an uncertainty of approximately 0.2 degrees
Celsius in each
of our measurement points. Given our measured average
attenuation length $\langle L_{atten}\rangle$, in order
to extract a
numerical estimate of the attenuation length as a function of depth
we must extrapolate the temperature at a depth of 97 m to the
temperature at the bedrock. To our knowledge, a full temperature profile
down to bedrock at our site is not currently available.
In the absence of a dedicated scan of the frequency dependence of
the received signal, we assume flat response of the transmitter
and receiver between 200 MHz and 1000 MHz, ignore any
frequency dependence of $L_{atten}$, and weight the received
signal amplitude by $\lambda$ to include the 
expected effective height dependence of the transmitter
and receiver antennas. Using bedrock
reflectivity equal to unity, and an average
attenuation length of 450 m,
a parabolic extrapolation of temperature
down to a warm bed gives an implied attenuation
length at -50 C and 380 MHz of 1480 m, 
within 2\% of the central value obtained at
South Pole\citep{Bar2005}.
A flat extrapolation at constant temperature gives a value
only half as large.

%\begin{figure} \centerline{\includegraphics[width=12cm]{matsuoka_oneway_atten.eps}} \caption{Origin is South Pole, units are km.} \label{attenmodel.eps} \end{figure}

\subsection*{${\rm S}_{12}$(ice) measurements - signal arrival time considerations}
\subsubsection*{Birefringence considerations} 
In principle, ground-penetrating
radar can be used to probe birefringent asymmetries, for
both the real and imaginary pieces of the dielectric constant. A
continuous wave (CW) network analyzer signal, fed into a Transverse
ElectroMagnetic (TEM) horn,
can be used to excite one particular polarization. In the case where
that signal polarization axis coincides with one of the linear birefringence
axes (aka ``optical axes'', or ``ordinary'' and
``extraordinary'' axes), no
birefringent asymmetry is observed. In the case where the 
signal polarization axis projects onto two orthogonal 
birefringence axes, 
a birefringent asymmetry ($\delta_{\epsilon'}\ne$0) in the real part of
the dielectric constant will result in interference between the two
orthogonal signals arriving at the receiver, with some frequency and
pathlength-dependent reduction in observed signal magnitude. 
%In general, therefore, the measured amplitude will be smaller than when the signal polarization axis is aligned along a birefringence axis. 
%http://www.agu.org/pubs/crossref/2003/2003JB002425.shtml; Matsuoka et al., JOURNAL OF GEOPHYSICAL RESEARCH, VOL. 108, NO. B10, 2499, doi:10.1029/2003JB002425, 2003
%igloo.gsfc.nasa/gov/wais/pastmeetings/PPT06/wais06sl.ppt
In the case where the signal polarization is at 45 degrees relative to
the birefringent axes, each of the equal-amplitude HH and VV
projections would
display a 
reduced amplitude over the time when only one projection is being
measured, followed by an
interference pattern with
amplitude varying from zero to the maximum observed for the unrotated
orientation, over
the timespan when the signals propagating along the birefringent axes
interfere at the receiver antenna, again followed by a reduced-amplitude
post-interference single-polarization signal. 
%Over the time interval for which both components contribute to the net received signal, the measured signal amplitude will vary as $\cos(\omega\t_1)+\cos(\omega\t_2)=2\cos(\omega<t>)\cos(\omega\delta t/2)$, with $<t>$ the averge propagation time and \delta t$ the propagation time difference between the two components. Figure \ref{fig:sim_biref.eps} models the physical implications for signal emission (monochromatic signal) as a function of the emission angle relative to the birefringence axes. \begin{figure} \centerline{\includegraphics[width=12cm]{sim_biref.eps}} \caption{Idealized model of birefringence, setting thermal noise to zero. Solid red and dashed green traces show signals received for polarizations oriented along the two birefringence axes, having different indices of refraction (n1 and n2, respectively). Dashed blue and dotted cyan curves model interference between the two polarizations, obtained when signal electric field vector lies between two birefringence axes. Blue dashed (cyan dotted) trace represents expected signal when horn is oriented at 45 degrees (22.5 degrees) with respect to  two birefringence axes. We assume a perfect receiver system. Note the time-domain offset modeled for the start time of the two signals.} \label{fig:sim_biref.eps} \end{figure}
Note that this gives us one criterion
for the antenna orientation which best matches the birefringent axes -- in
that configuration, the observed time duration of the signal is minimized
and the observed peak amplitude is maximized.

%In the case where the birefringent asymmetry is entirely in the absorptive (imaginary) part of the dielectric constant ($\delta_{\epsilon''}$=0), the variation in signal is likely to be less detectable. Figure \ref{fig:birefsim.eps} illustrates the expected signal variation for a continuous wave signal, as the signal polarization axis for a transmitter/receiver pair are rotated relative to the birefringence axes, as a function of the magnitude of the birefringent asymmetry. \begin{figure} \centerline{\includegraphics[width=12cm]{birefsim.eps}} \caption{Illustration of expected variation in received continuous wave (CW) signal strength, as a transmitter/receiver pair are rotated relative to major and minor birefringence axes. $\alpha$=0 corresponds to alignment along the major axis, for which the signal transit time is minimized.} \label{fig:birefsim.eps} \end{figure}

Considerable literature and experimentation have been devoted
to ice birefringence. 
Some of the relevant
measurements thus far are summarized in Table \ref{tab:birefsum}.
Here we restrict ourselves to only the most
recent experimental efforts in Antarctica. 
Frequency-domain experiments typically measure the power reflected
off the bedrock as a function of orientation angle of transmitter
and receiver. Variations in the power are typically the
combined result of anisotropies in bedrock scattering and also
birefringence. Multiple measurements at a variety of orientation
angles and frequencies allow one to separate, to a large
extent, the two contributions.
Data collected on Brunt Ice Shelf near Halley Station resulted in
an estimate of the minimum birefringent asymmetry
in the effective permittivity ($\delta_{\epsilon'}$)
of 0.14--0.52\%\citep{Doake2002}, depending on location. The same
experimental apparatus employed at George VI Ice Shelf yielded
values typically $3\times$ smaller.

\begin{figure} \centerline{\includegraphics[width=12cm]{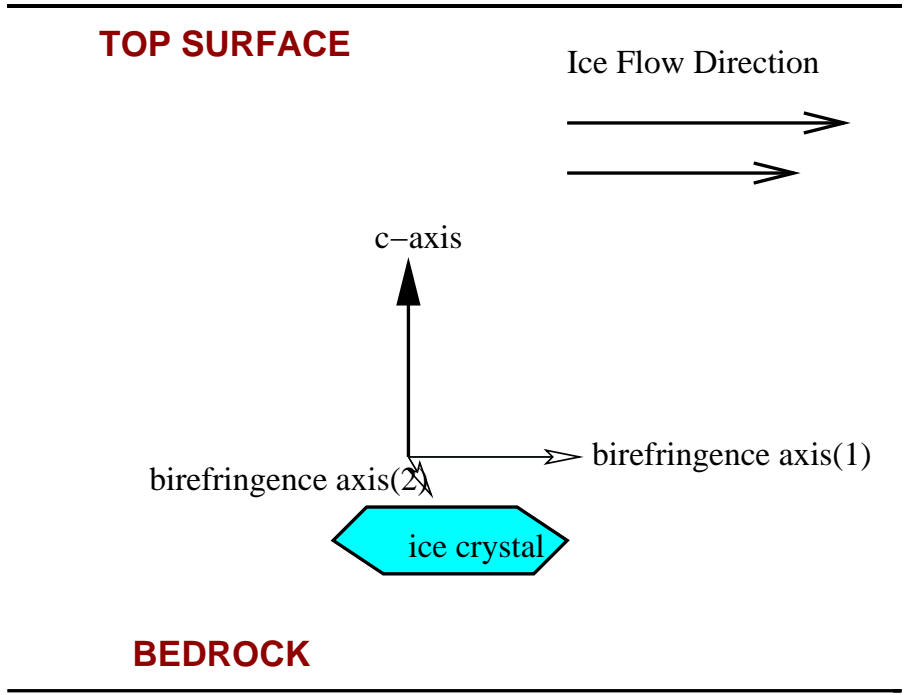}} \caption{Presumed orientation of birefringence axes, relative to ice flow.}\label{fig:biref-diagram.xfig.eps} 
\end{figure}

Continuous wave data taken using horn antennas 
rotated in the horizontal plane and
broadcasting along the axis perpendicular to the bed (``c-axis'')
over a 300-km traverse from Dome Fuji show a very clear 
sinusoidal pattern, with an amplitude modulation 
$\approx$80\% over $\pi$ radians\citep{Matsuoka-biref}. 
This is interpreted as direct observation of birefringence due to the
asymmetry in the crystal orientation fabric (``COF''),
caused by the ice flow itself along the gravitational
gradient and resulting in crystal orientation perpendicular
to the flow direction\citep{Rippin-2003}. Schematically, the geometry of the
crystal orientation, relative to the putative birefringence
axes, is shown in Figure \ref{fig:biref-diagram.xfig.eps}.

Direct measurements of Vostok ice cores, however,
%http://www.awi-bremerhaven.de/Publications/Ham2005b.pdf
indicate no preferred ice crystal orientation over the first 400 meters
in depth,
followed by a mixture of COF both parallel to,
as well as perpendicular to the c-axis (i.e., along
the vertical, or ${\hat z}$ and along an axis perpendicular to the 
horizontal ice-air interface) over the next 1400 m. Below that value,
the COF is found to align along the c-axis, with crystals preferentially
stretched along the vertical. 
We note also that measurements thus far indicate that sheet
ice has a uniform density below $\sim$200 m, so any alignment of the
internal fabric evidently
does not result in any bulk change in the crystal
packing.
%http://earthweb.ess.washington.edu/matsuoka/wais/slides_project_meeting_oct.swf

\message{compared with a value of 2 percent for single ice crystals - 
reference?}
\begin{table}[htpb]
\caption{Summary of birefringence measurements to date. \label{tab:birefsum}}
\begin{tabular}{c|c|c|c} 
Group & Locale & $\delta_{\epsilon'}$ Result & Comment \\ \hline
Matsuoka {\it et. al.} (1997)\cite{Matsuoka97} & Lab Ice & $\sim$3.4\% & 39 GHz \\
Fujita \& Mae (1993)\cite{Fujita-1993} & Lab Ice & $\sim$1.1\% & \\ 
Doake, Corr, Jenkins (2002)\cite{Doake2002} & Brunt Ice Shelf & $>$0.14--0.47\% & \\
Doake, Corr, Jenkins (2002)\cite{Doake2003} & George VI Ice Shelf & $>$0.05--0.15\% & \\
Woodruff \& Doake (1979)\cite{WoodruffDoake1979} & Bach Ice Shelf & 0.52\% & \\
Fujita {\it et. al.} (2003)\cite{Fujita-2003} & Mizuho Station & measurable & \\ 
Fujita {\it et. al.} (2006)\cite{Fujita-2006} & Dome Fuji & 0.05\% & $>$500 m, 
multi-parameter fit \\
Fujita {\it et. al.} (2006)\cite{Fujita-2006} & Mizuho & 1.5\%-3.5\% & $<$500 m, 
multi-parameter fit \\
\end{tabular}
\end{table}

As an alternative to frequency-domain measurements, one can directly
measure asymmetries in the time domain using a pulsed signal. This approach
has the disadvantage of obscuring the frequency dependence of the signal
propagation, but the advantage of isolating
through-air
signal leakage due to the side lobes of a 
typical horn beam pattern and/or other
possible multi-path effects. In this
case, $\delta_{\epsilon'}>0$ results in a measurable time
difference between two signal polarizations; 
$\delta_{\epsilon''}>0$ results in an amplitude difference between
two polarizations. 
%We caution that this measurement, like the CW measurement mentioned above, effectively probes only two of the three axes which are otherwise observable in direct ice-core analyses.

\subsubsection*{Experimental Comparison of Bottom Reflection signals}
\begin{figure} \centerline{\includegraphics[width=12cm]{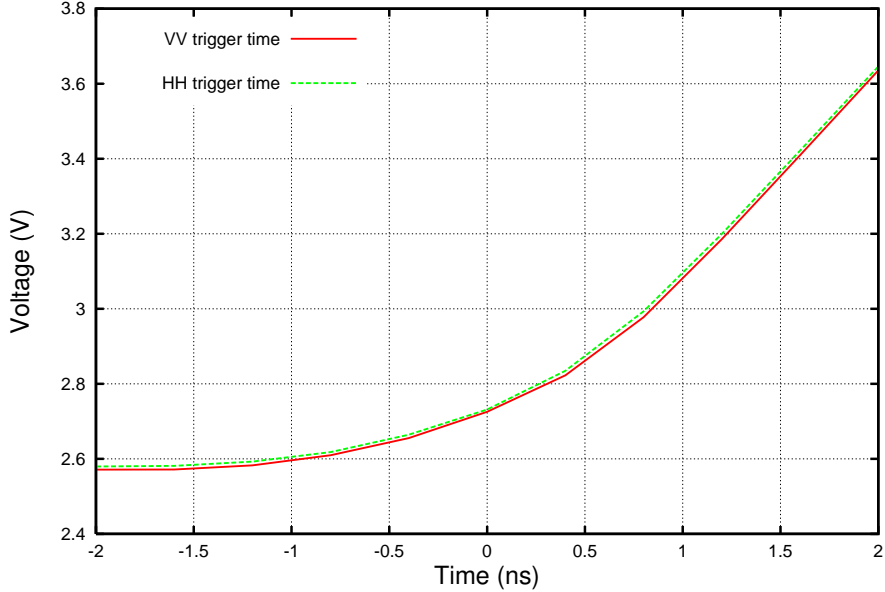}} \caption{Comparison
of trigger time for bottom reflection measurements
for HH vs. VV orientations (60 event average). 
Trigger threshold is set to 2.72 V. Time
difference in $t_0$ between HH vs. VV measurements is insignificant
compared to time delay measured between received signals.} 
\label{fig:TD-t0.eps} 
\end{figure}

Bottom reflection data were taken in ``VV'', ``HH'', and ``VH'' orientations. 
%As Figure \ref{fig:TD-HH-VV-t0.eps} shows, \begin{figure} \centerline{\includegraphics[width=12cm]{TD-HH-VV-0.eps}} \caption{Trigger time for bottom reflection measurements, various orientations. Threshold crossing occurs at t=0 seconds.} \label{fig:TD-HH-VV-t0.eps} \end{figure}
The start times ($t_0$) from the pulser %leakage propagating to scope thru air
 for our measurements
are found to be identical to within $\sim$200 ps. 
Figure \ref{fig:TD-t0.eps} compares the trigger crossing time for the ``VV'' and ``HH'' orientations. 
%We again note the presence of an apparently long tail in the reflected signal, as shown in Figure \ref{fig:HH-1165563686-full.eps}. 

\begin{figure} \centerline{\includegraphics[width=12cm]{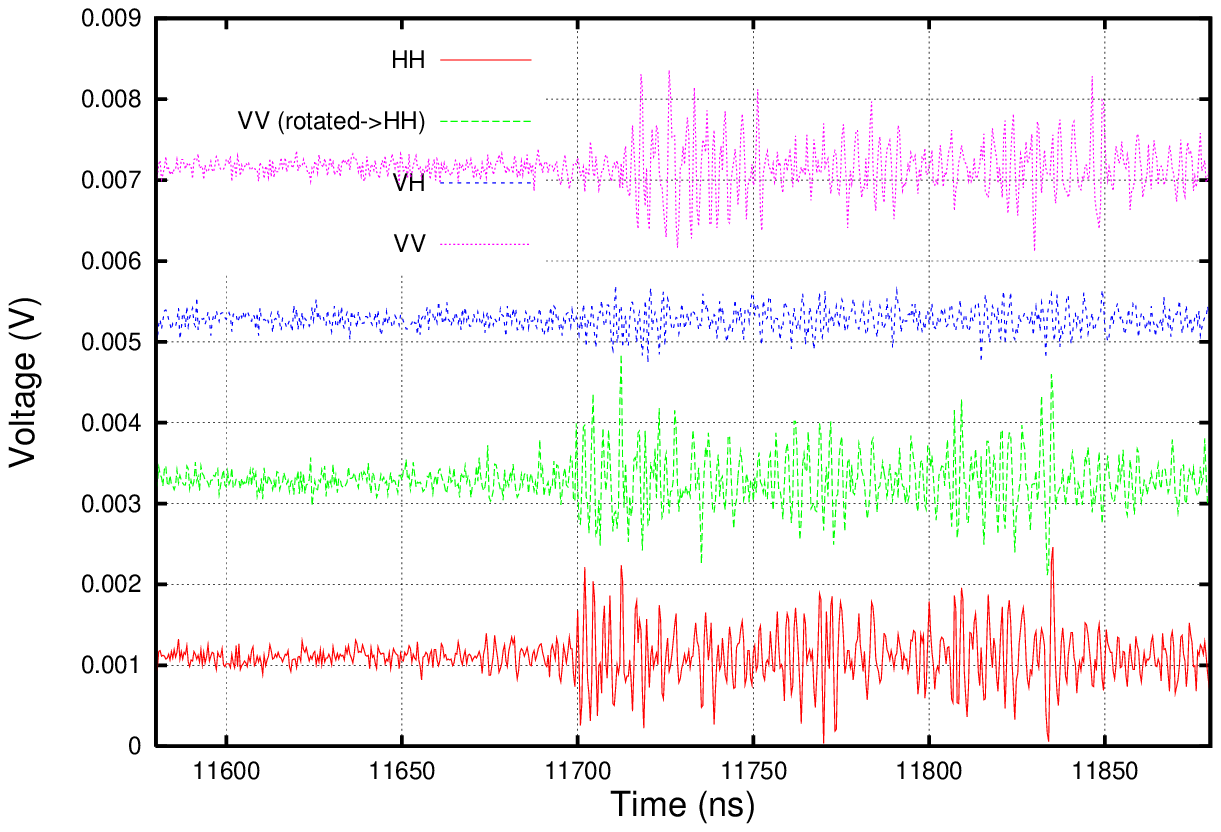}} \caption{Received signal as a function of time for indicated orientations.
``VV$\to$HH rotate'' refers to the Tx and Rx orientation for which
the VV axes of both Tx and Rx
have been rotated into the initial HH orientation.
In this (and successive) Figures, 
the three uppermost
signals have been vertically offset to 
enhance visual clarity.}
\label{fig:TD-HH-VV-1.eps} 
\end{figure}

\begin{figure} \centerline{\includegraphics[width=12cm]{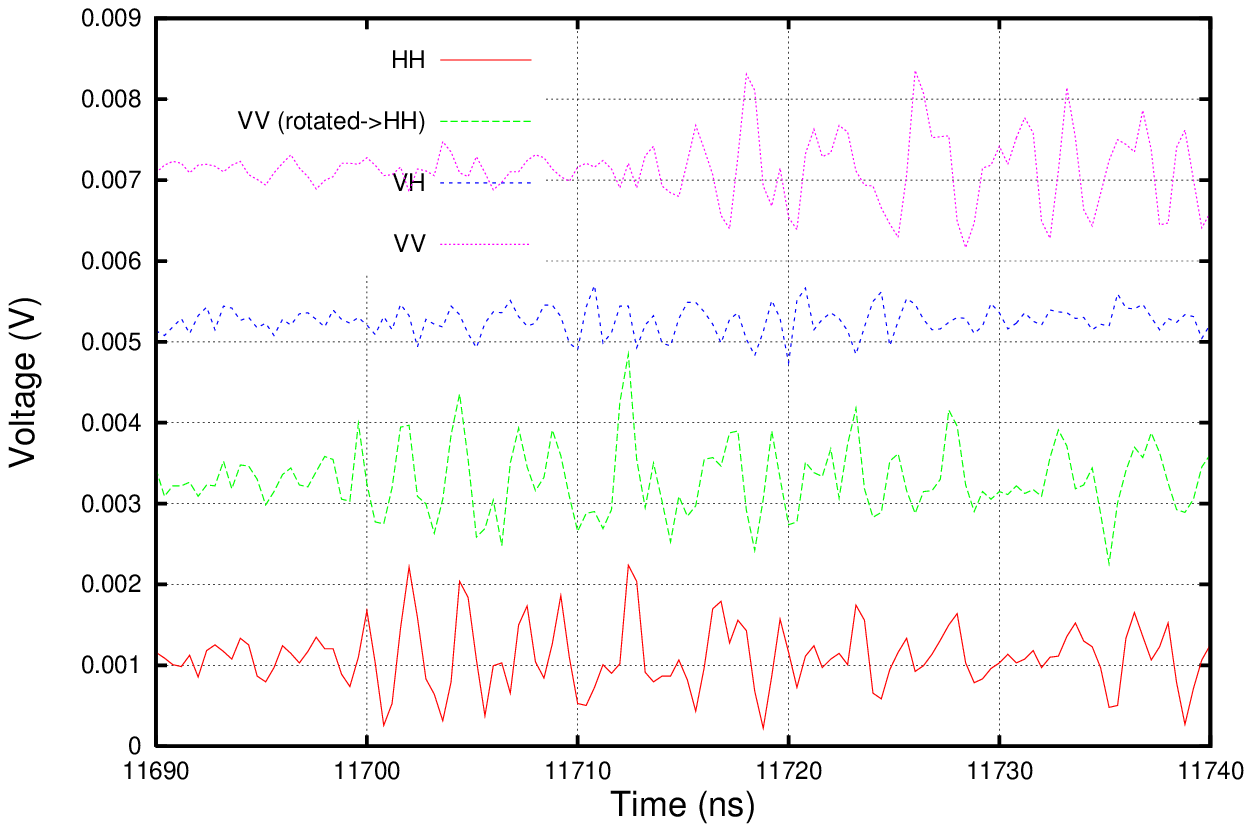}} \caption{Zoom of previous Figure.} \label{fig:TD-HH-VV-2.eps} 
\end{figure}

Comparing ``HH'' to
``VV'' reflections, we observe that
the echo time recorded for the ``HH'' signal is
advanced by 15 ns relative to the ``VV'' echo signal time (Figure \ref{fig:TD-HH-VV-1.eps}, and
zoomed in Figure \ref{fig:TD-HH-VV-2.eps}). 
Aside from the unlikely
possibility that the bottom surface has local `patches' which favor different polarizations, it would
appear that the most likely explanation for this observed time difference is due to wavespeed 
differences along two axes. 
Interpreted as birefringence, the implied asymmetry is approximately 0.12\%. %, considerably smaller than the estimate of 3.4\% made elsewhere (albeit at 39 GHz\citep{Matsuoka97}). 
Although not fully probed owing to time and cable length limitations, and
also problems
with data acquisition waveform capture (the voltage
resolution setting
on the digital oscilloscope was, unfortunately, unusably coarse),
we note that:
\begin{itemize}
\item the magnitude of the HH vs. VV signals were relatively constant
when the receiver antenna was
displaced along the N-S axis by approximately 170 m (in that case,
the illuminated reflecting
area of bedrock
should be displaced by $\sim$85 m) 
and then 
 rotated in 22.5 degree steps over 180 degrees. The intent here was to
probe the specular component of the surface scattering and attempt
to discern variations in peak receiver voltage, as a function of
orientation. The measured
values of peak receiver voltage, read off the
TDS694C oscilloscope screen,
are presented in Figure \ref{fig:draw_pol_phi_scan.eps}.
\begin{figure} \centerline{\includegraphics[width=12cm]{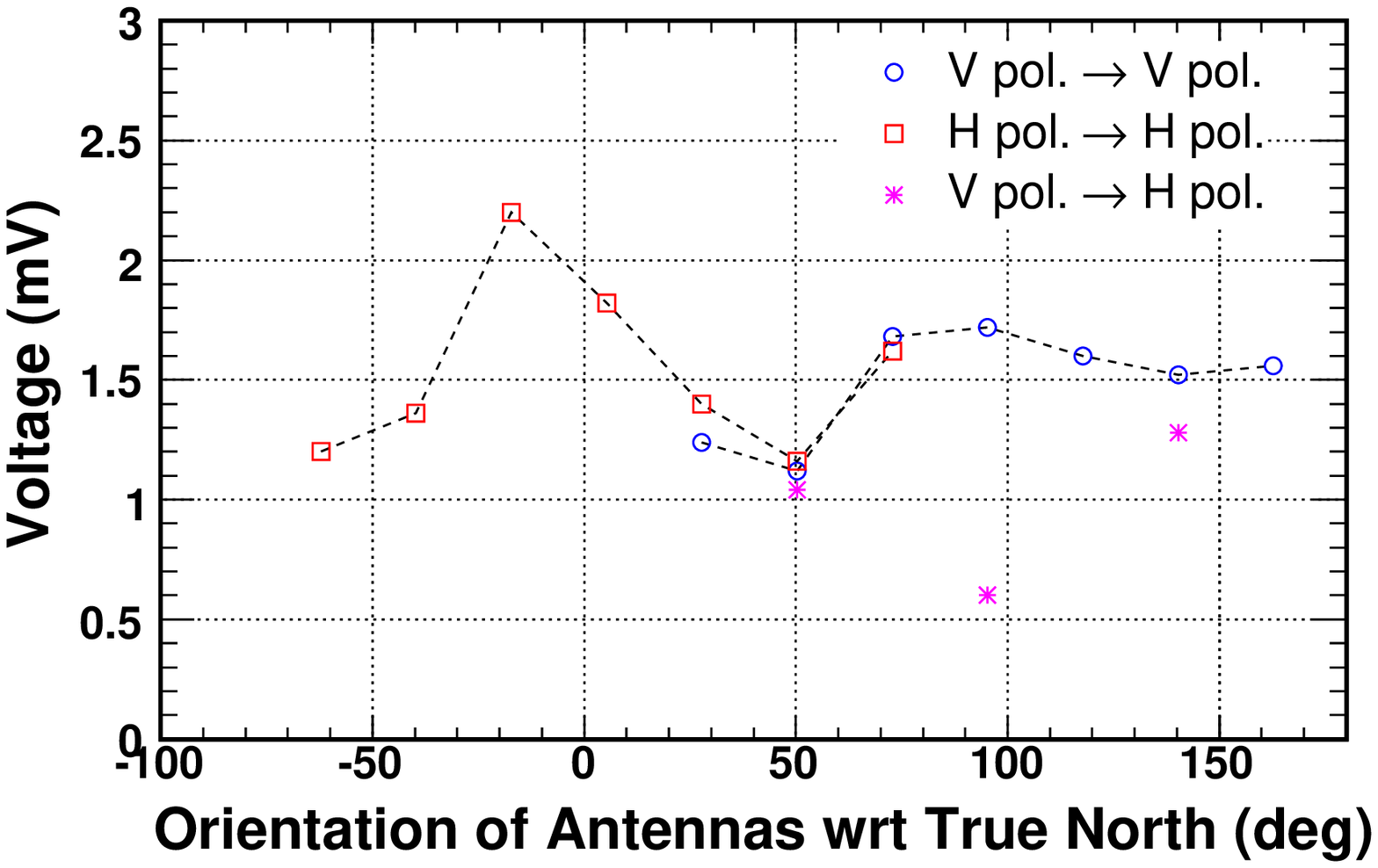}} \caption{Maximum measured bottom reflection voltage (as read directly from oscilloscope)
with surface receiver horn
displaced 170 m south of initial measurements. Typical measurement uncertainty is estimated to be 
of order 20\% at each point. ``Negative'' corresponds to a clockwise
rotation from true North; i.e., in the direction towards East.} \label{fig:draw_pol_phi_scan.eps} 
\end{figure}
We observe from Figure \ref{fig:draw_pol_phi_scan.eps} that:
a) our initial ``VV'' orientation of --14.8 degrees seems to be
close to the maximum voltage orientation; b) from interference effects,
and assuming that the birefringent asymmetry in the real part of the
dielectric constant is substantially larger than in the imaginary
component, we would naively expect maxima at intervals of $\pi/2$,
corresponding to those orientations for which the antenna polarization
axis is aligned with either of the optical axes;
c) absent physical effects which
rotate the polarization plane of the propagating signal, we
would expect the received cross-polarization ($V\to H$, e.g.)
fraction of the received signal to be approximately constant. 
The largeness of the point-to-point systematic errors 
and the lack of comprehensive data notwithstanding,
such constancy is not obviously observed. Rather, there appears to
be an anti-correlation between ``co-pol'' (VV or HH) signal strength and
cross-pol (VH) signal strength -- when the former is largest, the
latter is smallest, and vice versa.
\item To check antenna systematics, the surface horn
transmitter and surface horn receiver were rotated in the horizontal
plane by --90 degrees; in that configuration, 
we find HH(rotated)=VV(unrotated), and
VV(rotated)=HH(unrotated). This indicates that the observed time-domain
asymmetry is not an artifact of antenna effects.
\end{itemize}

\begin{figure} \centerline{\includegraphics[width=12cm]{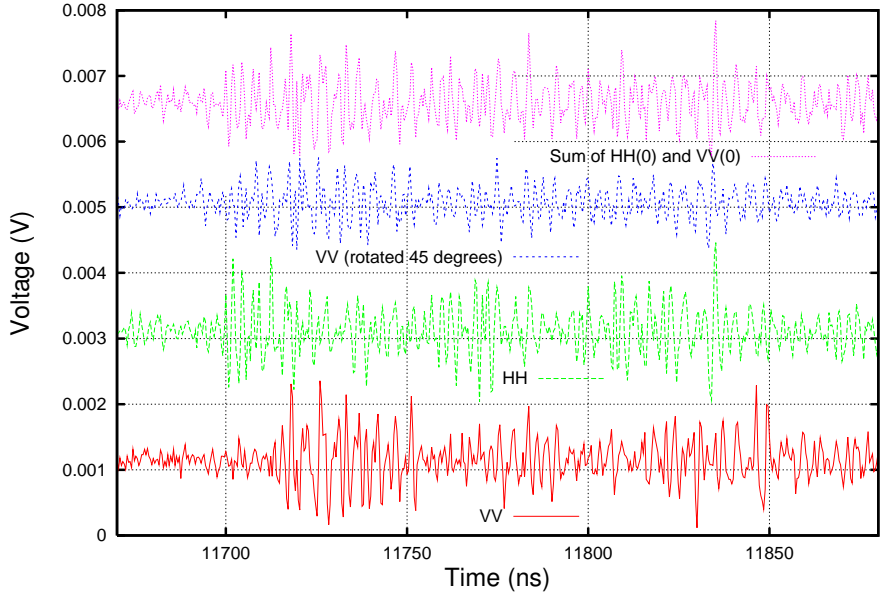}} \caption{Vector sum of HH and VV signals (our birefringence model, top) compared
with data received signals, as indicated. Plots have been vertically offset
for visual clarity. In the context of our model, the top plot should 
correspond to the second-from top plot (``VV (rotated 45 degrees)'').} 
\label{fig:TD7.eps} 
\end{figure}

Figure \ref{fig:TD7.eps} displays the received signals for both
the original VV and HH orientations, the signal
observed when the surface antennas are each rotated by --45 degrees 
($\pi$/4), and the vector sum of the VV and HH signals. The
latter is
obtained by projecting the VV and HH data onto an axis bisecting
VV- and HH-, and adding the
projected components. Neglecting thermal
noise contributions, which would add a random
voltage with rms 
$V^{rms}_{thermal~noise}$ point-to-point, our VV+HH sum
model should coincide with the --45 degree
data.
This comparison is, unfortunately, inconclusive.

%\begin{table}[htpb]\caption{Summary of measured peak voltages for various orientations, surface separation between Tx and Rx $\sim$100m (position 2). Measurement errors are estimated to be $\sim$10\%.}\begin{tabular}{c|c|c}Angle wrt True North (deg.) & $V_{peak}(HH)$ & $V_{peak}(VV)$ \\ \hline 0 & & \\ 22.5 & & \\ 45 & & \\ 67.5 & & \\ 90 & & \\ 112.5 & & \\ 135 & & \\ 157.5 & & \\ 180 & & \\ \end{tabular}\end{table}

\subsubsection*{Cross-Correlation Analysis}
\begin{figure} \centerline{\includegraphics[width=12cm]{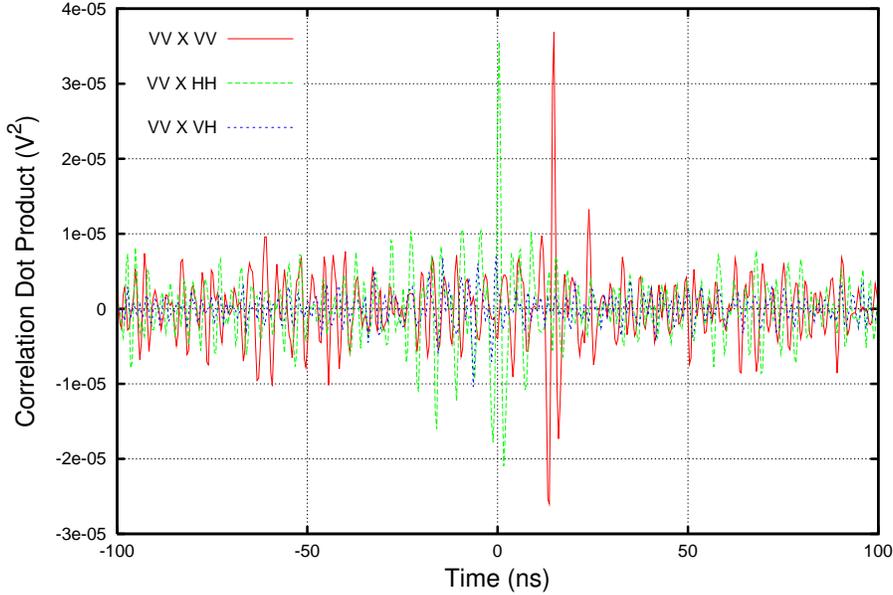}} \caption{Cross-correlation parameter, as described in text.} 
\label{fig:TD-xcor.eps} 
\end{figure}

A cross-correlation analysis can also be used to both extract the
signal hit-time, and also assess the degree of `similarity' of the
recorded waveforms. In this analysis, a 200 ns-long segment of
the waveform, centered on one of
the reflection signals observed in the initial VV-orientation,
comprises the `filter' pattern. The `VV$\to$HH rotated', HH and
HV waveforms comprise `templates'. Over a 4000 ns template
segment, centered on the observed bottom reflection in each
of the template waveforms, we calculate the correlation parameter
${\cal C}=\Sigma_{i'=0}^{3800}\Sigma_{i=0}^{200}V_{i,filter}V_{i',template}$.
Figure \ref{fig:TD-xcor.eps} illustrates the similarity in the HH
and VV waveforms, which have nearly identical cross-correlation
maximum amplitudes.

\section*{Faraday Rotation}
An apparent propagation asymmetry can also be due to the interaction of
an electromagnetic wave with an external magnetic field (Faraday rotation).
The interaction of electromagnetic radiation having a propagation
wave vector (${\vec k}-$) parallel to an external magnetic field ${\vec B}$
can lead to an apparent rotation of the 
electric field polarization plane. This can be
visualized as a decomposition of the planar wave into left-
and right-circular polarization components, which interact differently
with the ambient field and propagate at different
velocities. In that sense, Faraday rotation can
be thought of as birefringence of circular, rather than
linearly polarized basis states. The degree of angular
rotation $\beta$ over a distance $d$ is specified by the
Verdet constant (given by ${\cal V}=-(e\lambda/2mc)dn/d\lambda$) via
$\beta={\cal V}Bd$. 
For water at 20$^\circ$ C, the measured value of ${\cal V}$ is
0.00038 rad/m/Gauss at optical
wavelengths, where it has been studied most extensively. 
%Hecht: (1.31 minutes/(60 min/degree))*(3.14159/180)
Although the value of the local magnetic field was not measured at our 
particular site, the proximity to the South Magnetic Pole
%http://www.dsri.dk/multimagsatellites/virtual_talks/desantis/images/slide07_trans.jpg
(defined as the point where the geomagnetic field is vertical) at
$65^\circ$S and $139^\circ$E implies that the magnetic field is
largely aligned along the ${\hat z}$, or vertical ($c$-axis).
Using Verdet constant data for water (this may not be directly
applicable to ice, however, ice data were not available) and
a value for the earth's field of 1 Gauss leads to
an expected rotation of $\theta_{FR}\sim$20 degrees over the one-way path from
surface to bedrock. Since the underlying bedrock has a larger 
index-of-refraction than the ice sheet above it, the electric field 
vector is expected to invert (i.e., rotate by 180 degrees around
${\vec k-}$) upon reflection, after which the electric field
vector rotates in the opposite sense, as viewed from behind.
The two effects add, so that the total rotation, relative to the
sent signal, will be $\theta_{tot}=2\theta_{FR}+\pi$, and could
therefore affect the measured leakage into the `cross-polarized'
signal. However,
Faraday rotation is not expected to give a net time stagger between
the two measured linear polarizations.

%http://en.wikipedia.org/wiki/Faraday_effect: A positive Verdet constant corresponds to L-rotation (anticlockwise) when the direction of propagation is parallel to the magnetic field and to R-rotation (clockwise) when the direction of propagation is anti-parallel. Thus, if a ray of light is passed through a material and reflected back through it, the rotation double.s

%http://www.indiana.edu/~p451f06/Faraday.pdf, and references therein: esp. Landolt-Biornstein, Physics Data Tables, 6th edition, QC61.125, Vol. I/1, p. 406 and G.S. Monk, Light: Principles and... QC357.M7
%http://www.lsbu.ac.uk/water/microwave.html gives plots of n(omega).
%The relative permittivity varies with wavelngth as: $\epsilon_r'=({\epsilon_S-\epsilon_\infty})/(1+(\lambda_S/\lambda)^2)$. In this expression, $\epsilon_S$ is the dielectric constant in the `static'low frequency limit ($\sim$99 for Ice Ih at --20 C); %www.lsbu.ac.uk/water/data.html $\epsilon_\infty$ is the dielectric constant in the high-frequency limit. With $\lambda\sim$1 m...

\section*{Broadcast through firn}
\begin{figure} \centerline{\includegraphics[width=12cm]{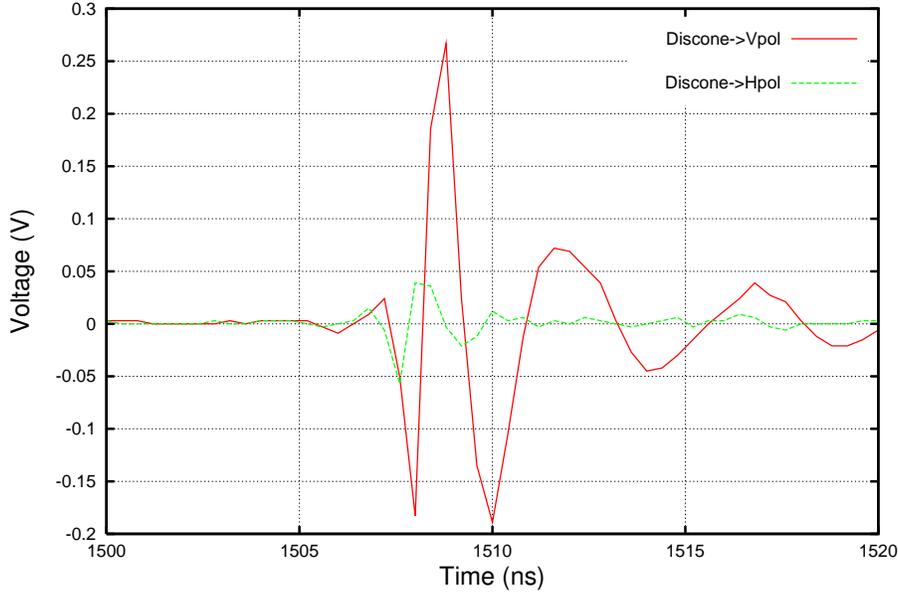}} \caption{Comparison of typical Vpol signals measured by surface horn receiver from in-ice discone antenna.} 
\label{fig:TD-discone-2-V-vs-H.eps} 
\end{figure}

Our measurement of the birefringent asymmetry is, of course, only
the asymmetry projected onto the vertical signal propagation axis.
To probe possible asymmetries in the horizontal
plane (and also surface transmission effects), 
narrow pulses were broadcast from the discone antenna buried 
at a depth of 97 m into the ice, and the signal measured in a
surface horn receiver antenna. 
Due to the small calibre of the
borehole,
only a vertically polarized antenna
(dipole or discone) with a large vertical:horizontal aspect ratio could
be used as a buried transmitter. Figure \ref{fig:TD-discone-2-V-vs-H.eps}
displays the Vertical-polarization 
and Horizontal-polarization
signals received by the 
surface horn when broadcasting from the
in-ice, vertically-oriented discone. 
The observed Hpol signal is consistent with the
known cross-talk and isolation 
between the vertical and horizontal polarizations of the surface horn
receiver. Given the short in-ice pathlength, and cross-talk
complications, it is somewhat difficult to
interpret the equality of the observed signal times
in terms
of birefringence. Moreover, since most of the signal path is through the
firn, it is quite possible that the ice crystal fabric has not yet, at this limited
depth, established a preferred orientation.
Taken at face value, the observed asymmetry of 0.12\% in our $S_{12}$(ice)
measurements would imply a difference in
the Hpol vs. Vpol signal arrival times of about 700 ps, just at the
edge of our sensitivity. 
%Nevertheless, visually, we note an apparent sub-ns offset in the two received waveforms.

\begin{figure} \centerline{\includegraphics[width=12cm]{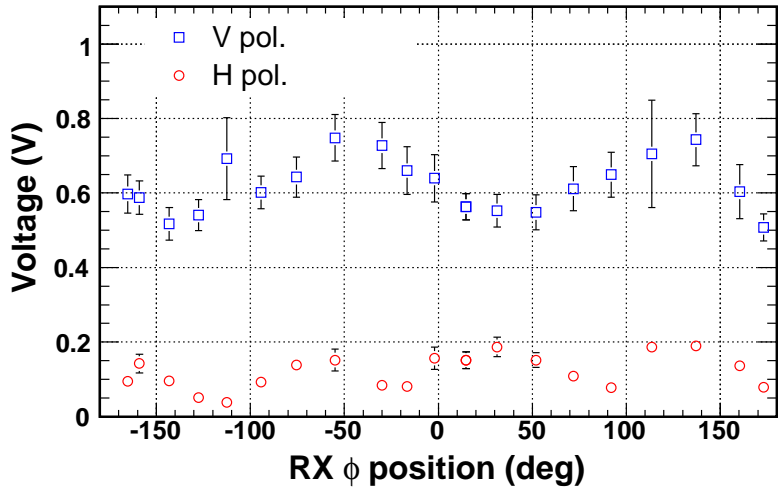}} \caption{Received surface horn signal strength, broadcasting from in-ice discone, as a function of azimuthal angle.} \label{fig:draw_phi_scan.eps} 
\end{figure}

\begin{figure} \centerline{\includegraphics[width=12cm]{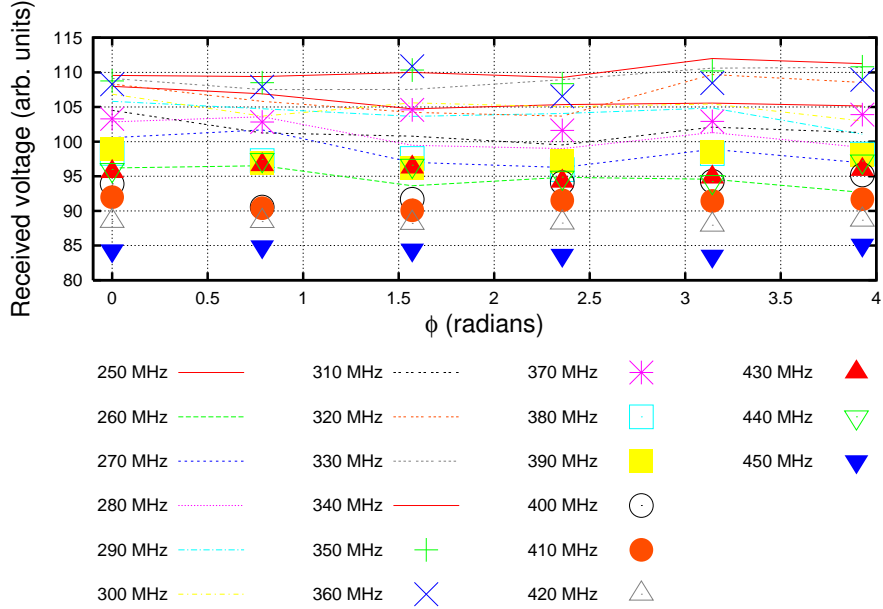}} \caption{Received azimuthal power, broadcasting with discone (in-air) to 
surface
Seavey horn, as a function of frequency.} \label{fig:discone2horn-phi.eps} 
\end{figure}

The peak received signal strength (corrected for the distance
from discone to receiver) shows a clear sinusoidal dependence, as a 
function of azimuth, as shown in Figure \ref{fig:draw_phi_scan.eps}.
We have investigated the possibility that the observed azimuthal signal
strength dependence might be a simple consequence of the intrinsic beam pattern
of the discone itself. A second discone, similar in construction to the
in-ice discone (the original discone was left in the ice for
the primary ANITA mission) was rotated azimuthally
in-air at an elevation of $\sim$2 m and the signal strength
measured in the surface receiver horn
(Figure \ref{fig:discone2horn-phi.eps}). The lack of azimuthal
variation observed over $\pi$ radians indicates that the
structure observed when broadcasting from the in-ice discone to the
surface receiver is not a consequence of the discone beam pattern.
The fact that the received signal magnitude seems
correlated with the local surface gradient suggests that
the observed variation may be due to the polar angle dependence of the
signal emission and reception. In that case, however, we expect a $2\pi$ rather than
a $4\pi$ modulation of the observed receiver signal. 

\begin{figure} \centerline{\includegraphics[width=12cm]{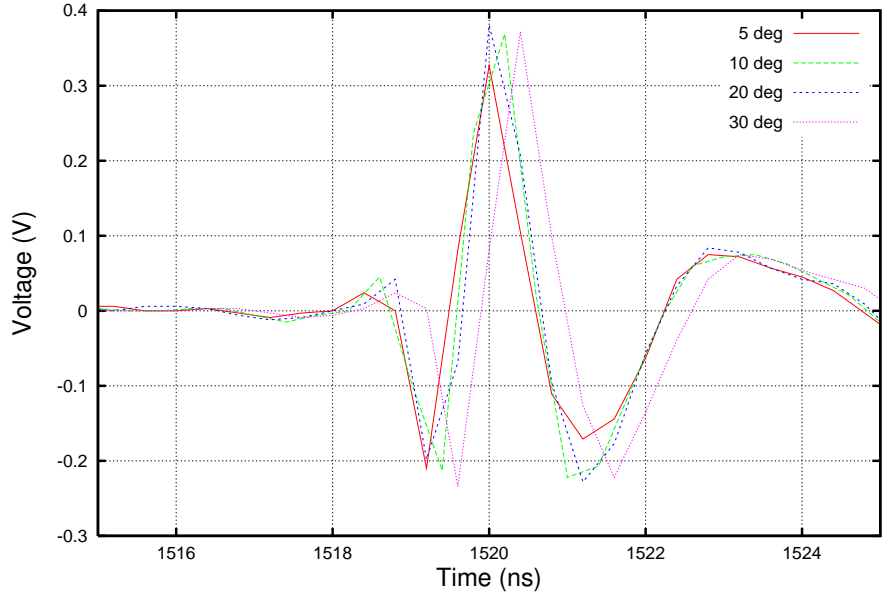}} \caption{Signal dependence (broadcasting from in-ice discone) on surface horn receiver cant angle; r=70 m.} \label{fig:TD-signal-V-theta-angle.eps} 
\end{figure}

Nevertheless, exit angle dependences were explicitly investigated.
By default,
the receiver horn antenna was oriented
with an inclination angle of $-11\pm 1.5$ degrees relative to the horizon
(i.e., pointing down relative to true horizontal). The local surface
elevation gradient resulted in a measured maximum difference
in snow exit angle of amplitude 
4 degrees. To investigate the polar angle dependence, the receiver signal was 
measured as a function of cant angle with respect to the horizontal 
(Figure \ref{fig:TD-signal-V-theta-angle.eps}).
The magnitude of the observed variation indicates that no more than $\sim$10\% of
the observed azimuthal variation might be due to local exit angle effects.

\subsection*{Surface Roughness Effects}
\begin{figure} \centerline{\includegraphics[width=12cm]{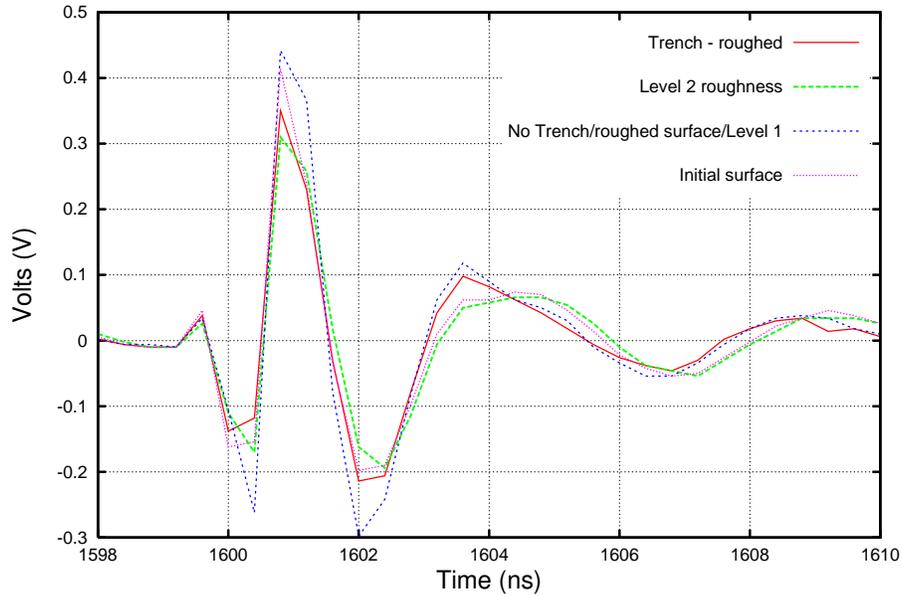}} \caption{Comparison of signals received for differing amounts of (artificially created) surface roughness in the region in front of the surface horn. For all these measurements, the surface receiver horn is elevated approximately 0.5 m, so that the center of the antenna is elevated approximately 1 m above the surface.} \label{fig:TD-roughness.eps} 
\end{figure}

The observed variation may also be the result of surface roughness
effects. We have measured the rms variation of the local snow elevation,
and the dependence of the observed signal strength on variations in
surface roughness. At a radial distance of r=70 m from the borehole, the
received signal strength was measured initially, then
the snow surface was `roughed' with a shovel.
Next, a shallow trench ($\sim$25 cm deep) was dug and `roughed'
again and the
signal strength re-measured.
Figure \ref{fig:TD-roughness.eps}
shows the results of this procedure. 
%Figure \ref{fig:draw_surface_roughness.eps}shows the distribution of surface measurements, \begin{figure} \centerline{\includegraphics[width=12cm]{draw_surface_roughness.eps}} \caption{Distribution of surface depths in artificially `roughed' area in front of horn, sampled over 2 m x 2 m area.} \label{fig:draw_surface_roughness.eps} \end{figure}
Table \ref{t:surf-roughness} shows the average
value of snow `depth' and also the rms-variation in the snow depth, 
sampled randomly in the 2m x 2m
patch of snow surface directly in front of the antenna. 
\begin{table}[htpb]
\caption{Summary of direct measurements of snow surface
depth variation. d=0 corresponds to the estimated average surface
level of the surrounding snow. r is the radial distance
between the surface horn receiver and the in-ice discone,
\=d gives the average of the 
measured snow depth values, and $\sigma_d$ is the rms of the
measured depth values. \label{t:surf-roughness}}
\begin{tabular}{ccccc} \\
r  & Samples 	& Surface condition & \=d & $\sigma_d$ \\ \hline
71 m & 26  & $1^{st}$-level roughness & 27.15 cm & 7.2 cm \\
71 m & 31  & $2^{nd}$-level roughness & 30.5 cm & 7.9 cm \\
71 m & 27  & in-trench & 34.6 cm & 6.7 cm \\ \hline
\end{tabular}
\end{table}
Given the fact that the 
surface roughness was considerably larger 
than that expected `naturally', we assess
the systematic uncertainty
in ANITA signal reception
due to surface roughness effects, based on our
measurements alone,
as $\sim$10\%. It should
be mentioned that we have conducted our measurements only
over an extremely limited area compared to the variety
of surface features actually probed by ANITA and this
estimate is not necessarily applicable to the entire
continent. 
%Quantitatively, the illuminated surface region relevant for our measurements is of order 1--2 meters, while for ANITA, the illuminated surface region is of order 30 m.

\begin{figure} \centerline{\includegraphics[width=12cm]{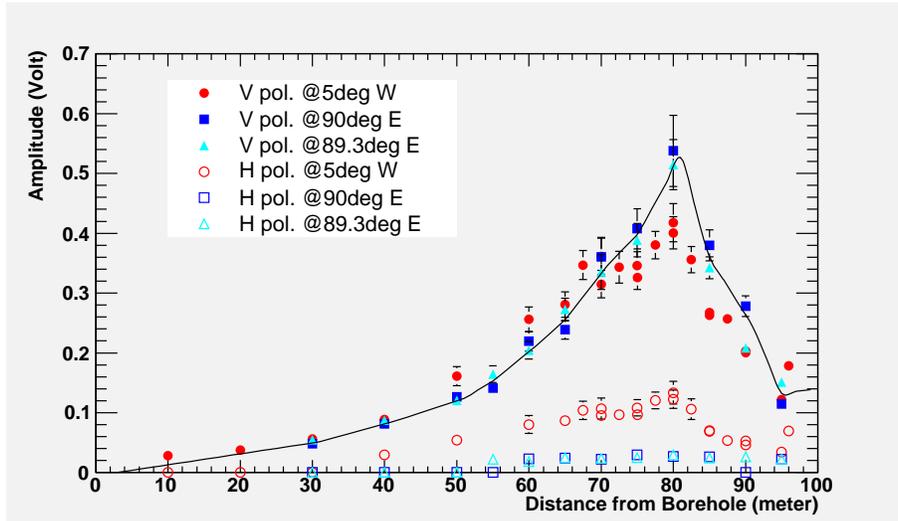}} \caption{Measured peak voltage in surface receiver horn (in-ice discone transmission) along three lines in azimuth, two of which are close to each other. 
Since the discone broadcasts over all polar angles, there is always a ray connecting the in-ice discone to the above-ice surface horn receiver.}\label{fig:draw_r_scan.eps} 
\end{figure}

As an additional measure of surface roughness effects, the transmitted
discone signal was measured by the surface horn receiver
along a constant azimuth, and also with
the surface horn at the same value of radial
distance but displaced in azimuth by 
approximately 1 m to the West, corresponding to the
spatial scale of the surface illuminated region. 
This displacement is sufficiently
small that there should be no significant change in received voltage
due to beam pattern uncertainties; we attribute any observed
variation to the different air-ice interfaces probed from the two
receive points. In this case, we again observe
$\sim$10\% variation in received signal amplitude.
Figure \ref{fig:draw_r_scan.eps} shows the signal strength as a function of radial distance from the borehole, along two slightly different lines in azimuth. 
The degree of scatter, for points very nearby in $\phi$ provides a measure of what we have not only attributed to 'surface roughness' systematic error, but also the single-shot measurement error. 

\subsubsection*{Amplitude dependence on density profile}
\begin{figure} \centerline{\includegraphics[width=12cm]{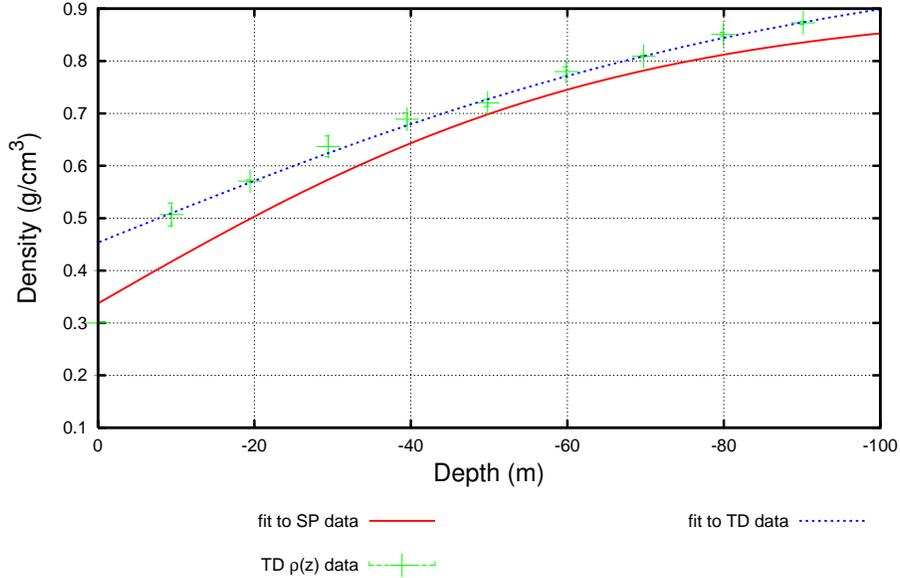}} \caption{Taylor Dome density data (Fitzpatrick, 1994, and Dixon, 2007) points and fit to density profile, compared with fit to density profile data at South Pole.
These profiles are used to calculate the expected propagation
between the in-ice discone and the surface receiver.}\label{fig:TD-rhovsz-100m.eps} 
\end{figure}

\begin{figure} \centerline{\includegraphics[width=12cm]{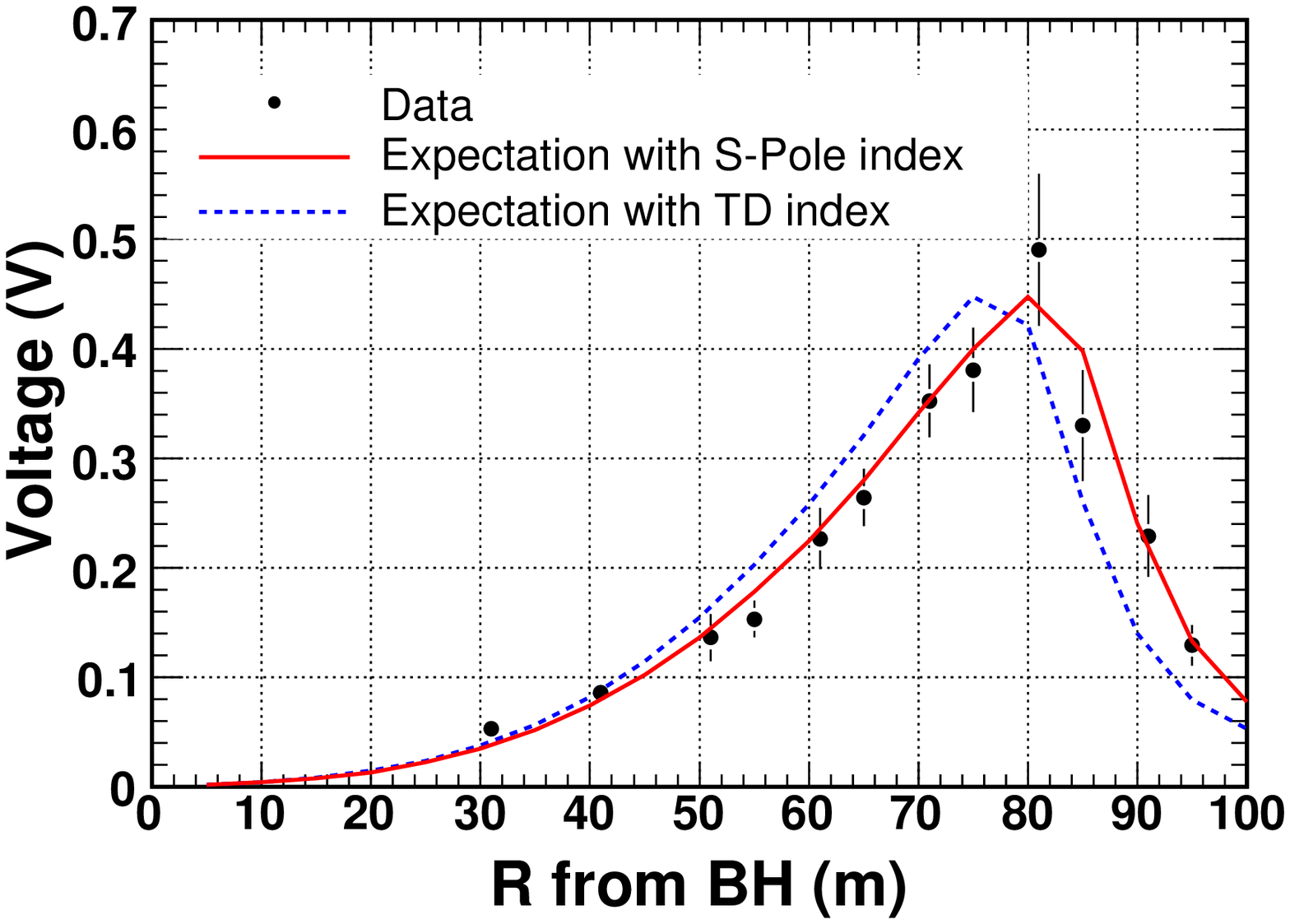}} \caption{Average of $90^\circ$ and $0^\circ$ measured peak voltages in surface receiver horn (in-ice discone transmission) with fit to expected amplitude, as a function of radius.}\label{fig:rscan2}
\end{figure}

In addition to the known discone and receiver horn beam patterns, and the
separation between them,
the signal amplitude at the surface receiver horn is primarily dependent on the
Fresnel coefficient at the ice-air interface. The incident scattering
angle as the ray emerges into the air therefore depends on 
the density profile $\rho(z)$, which determines the ray tracing through
the firn. In Figure \ref{fig:rscan2}, we show the result of a simulation
based on two test firn density profiles
 (Figure \ref{fig:TD-rhovsz-100m.eps},
showing only the first 100 meters of depth near Taylor Dome, from
Figure \ref{fig:TD-rhovsz-tobed.eps}, as well as the fit using South
Pole density measurement data overlaid).
These amplitude data slightly favor the South
Pole density profile as a better indicator of the correct firn density.

\subsection*{Signal arrival time dependence on density profile.}
\begin{figure} \centerline{\includegraphics[width=12cm]{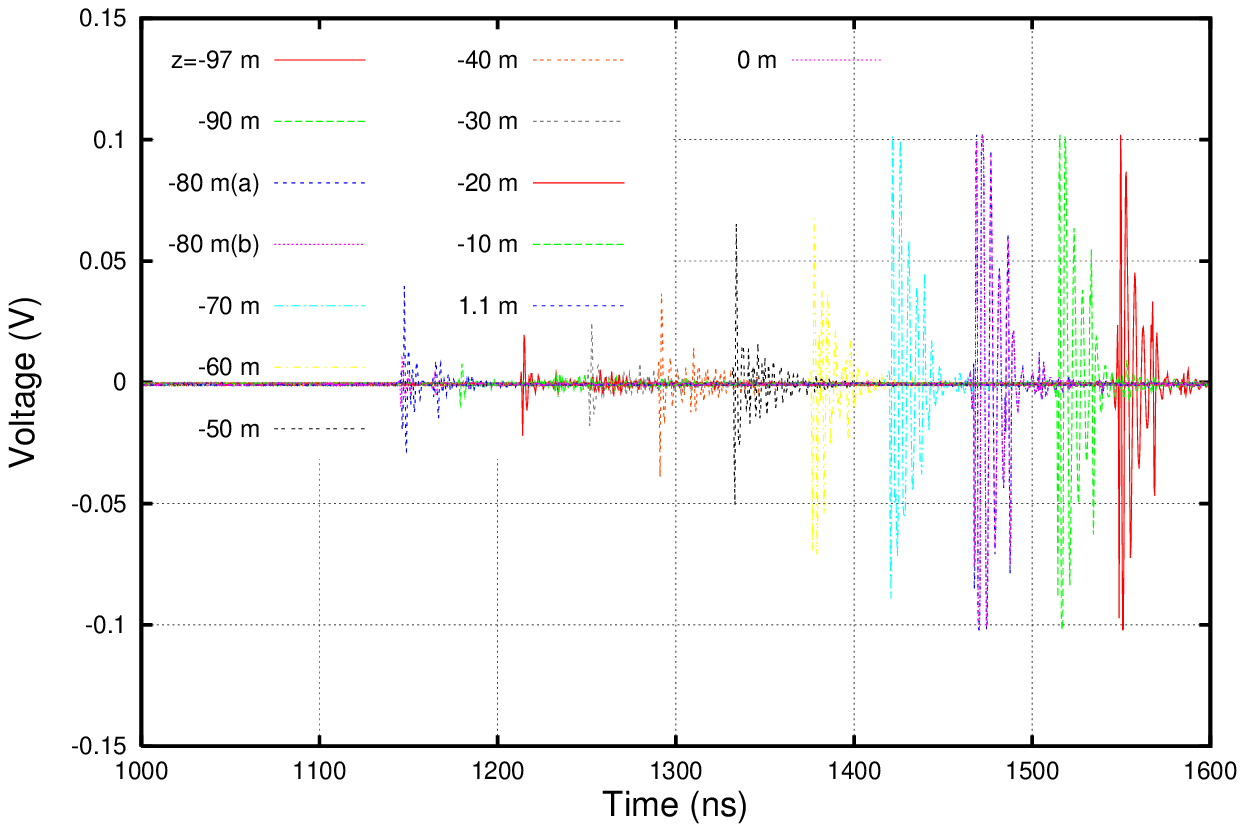}} \caption{Signals measured in 
surface horn as a function of depth of the transmitting discone; 
radial displacement between Rx and Tx is 70 m.
Voltage saturation occurs at $\sim$0.11 V (due to limited dynamic range setting of scope), as evident from 
the Figure.} \label{fig:discone2surfCV.eps} 
\end{figure}

\begin{figure} \centerline{\includegraphics[width=12cm]{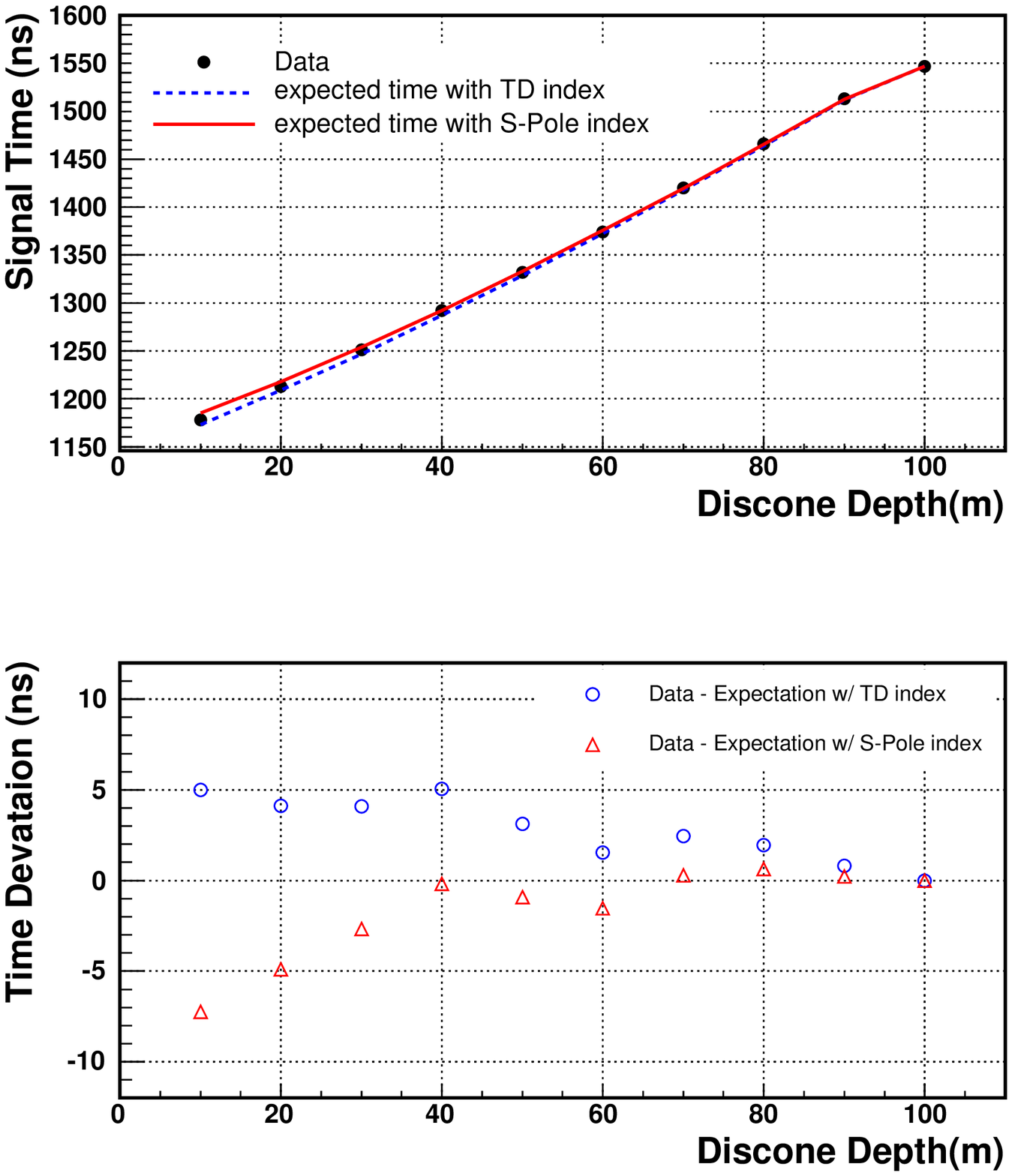}} \caption{Comparison of expected vs. measured hit times, assuming
index of refraction is related directly to density via Schytt equation.}
\label{fig:t_depth_with_deviation.eps} 
\end{figure}

In principle, the discone$\to$surface receiver
signal arrival time data
can also be used to directly verify the density profile over
the first 100 m of ice, as done previously at the South Pole\citep{Kr2005}. 
Figure \ref{fig:discone2surfCV.eps} shows the signals received as the
discone was lowered, at approximately 10 m intervals into the borehole.
Table \ref{t:dcone} 
gives the signal arrival time and amplitude characteristics (due
to operator error, the maximum possible measured voltage was only 0.1 V).
\begin{table}[htpb]
\begin{tabular}{c|c|c|c|c}
Depth & Recorded Signal time (ns) & Peak Voltage (V) & TD $\rho(z)$ fit time (ns) & SP $\rho(z)$ fit time (ns) \\ \hline
+1.1 m & 1146 & 0.040 & \\
0 m & 1145 & 0.010 &  \\
--10 m & 1178 & 0.009 & 1178 & 1178 \\
--20 m & 1213 & 0.020 & 1213.7 & 1210.6 \\
--30 m & 1251 & 0.024 & 1251.8 & 1246.3 \\
--40 m & 1292 & 0.040 & 1291.8 & 1284.7 \\
--50 m & 1332 & 0.064 & 1333.7 & 1325.4 \\
--60 m & 1374 & 0.071 & 1377.3 & 1368.1 \\
--70 m & 1420 & $\ge$0.1 & 1422.4 & 1412.2 \\
--80 m & 1466 & $\ge$0.1 & 1468.8 & 1458.5 \\
--90 m & 1513 & $\ge$0.1 & 1517.6 & 1506.2 \\
--97 m & 1547 & $\ge$0.1 & 1551.9 & 1540.0 \\ \hline
\end{tabular}
\caption{Comparison of measured signal arrival time, as a function
of depth, with signal arrival time predicted using the Taylor Dome
density profile used herein, and also the measured
South Pole density profile through the firn. Indicated times include
cable propagation times, in addition to transit times through
ice/air. \label{t:dcone}}
\end{table}
Assuming the index-of-refraction is related to density
via the Schytt model\citep{Schytt58} ($n(z)=1.+0.86\rho(z)$), we find
reasonably good agreement between our tabulated results
and the data taken from Taylor Dome itself. 
For comparison, we have also included the
expected times using the $\rho(z)$ profile using density
data from the South Pole.
Our propagation time measurements suggest that the
density profile at our experimental site is slightly
less sharp than the density profile obtained at Taylor Dome itself.
Figure \ref{fig:t_depth_with_deviation.eps} presents our results
graphically. In this case, rather than taking the ``zero'' hit time
as the time recorded when the discone was at a depth of 10 meters, we
compare our measurements to the 100 meter depth propagation time data.

%\subsection*{Comment on signal-strength dependence on radial distance (r)}

%By tracing rays through the firn, and determining the exit point of rays at the surface, we can use these measured values to estimate the sensitivity of the ANITA gondola receivers. The expected exit points are as indicated in Table \ref{t:exitpoints}.\begin{table}[htpb] \begin{tabular}{ccc} \\ $r_{ANITA}$ & $z_{ANITA}$ & $r_{exit,surface}$ \\ \hline 400 km & 30 km & 78.12 m \\ 400 km & 20 km & 78.34 m \\ 200 km & 30 km & 76.97 m \\ 200 km & 40 km & 75.84 m \\ \end{tabular} \end{table} Noting that the peak measured surface signal strength occurs at approximately r=80m, corresponding to a total distance of 125 m from the in-ice discone (and neglecting losses in the surface horn receiver which are not present for the on-board antennas), we can calculate the expected signal on the ANITA balloon, at a distance R from the in-ice discone to be of order (500 mV)$\times$(125 m/R). Assuming a threshold a factor of three times larger than the rms thermal noise voltage in the $\sim$1 GHz bandwidth of the ANITA receiver system ($3-\sigma_{kT,rms}(V)\sim 50\mu$V, the observable range of the ANITA receivers should be well beyond the physical horizon of $\sim$600 km. 

\section*{Conclusions}
Our initial goal was to measure the attenuation length of ice 
at a site close to the Antarctic coast (Taylor Dome) and
to compare that value with previous measurements at South Pole. This is one of
a series of anticipated experiments to map out ice properties over the large
area probed by the balloon-borne ANITA neutrino detection experiment.
Using a variety of somewhat redundant measurements, we estimate the
uncertainty in the expected ANITA signal strength due to 
uncertainties in transmission across the air/ice interface to be of
order 10\%.

We observe a clear signal,
reflected off of the underlying bedrock.
Determination of the exact signal strength for this bottom reflection
is complicated by uncertainties regarding the bottom topography and
the bedrock-ice reflection coefficient, as well as possible 
interference effects. 
For a variety of orientations, with both
Rx and Tx identically polarized, and probing a variety of
surface Tx/Rx locations over a 100-meter surface radius region,
we measure peak reflected voltage values that
are consistent to within approximately 30\%, with
a S:N of $\sim$4.0, implying attenuation lengths roughly 
consistent with measurements made at the South Pole.
%Our numerical estimates have generally erred on the side of yielding a conservative estimate of the attenuation length. More exact estimates of the reflection coefficient, as well as more sophisticated integrations of the reflected signal will likely add power to our estimate of the total received signal strength. 
Additionally, we find strong evidence for an asymmetry in
the real part of the ice dielectric constant, which we interpret as
the first observation of time-domain
birefringence. 

Due to the reduced maximum amplitude, the efficiency for registration
of ultra-high energy neutrinos, based on
a voltage threshold-crossing event trigger, may be
reduced if the birefringent asymmetry is 0.12\% and the result
of crystal orientation fabric (COF) alignment.
The quantitative impact of any possible
birefringence will depend on the experimental
details of a given data acquisition system and the location-specific
ice properties. 
Using a Monte Carlo simulation for the RICE\citep{Kr2006} 
experiment at the South Pole, consisting of 16 vertically-oriented
dipoles confined to a 200 m x 200 m x 200 m cube centered 200 m
below the surface at the South Pole,
we estimate a loss of $\sim$15\% in effective 
neutrino detection volume, roughly independent of neutrino
energy.
For the ANITA experiment, the effect is
expected to be considerably smaller, since the data acquisition
system integrates $\sim$8 ns after the
initial trigger signal, and
also since almost all the detected
neutrino flux corresponds to the case where the Cherenkov
electric-field vector is polarized predominantly in the
vertical plane, whereas birefringence should be most
noticeable for the case where the ${\vec E}$ polarization
vector coincides with one of the principal axes of the 
ice crystal (assumed to be oriented horizontally, in the
case where the fabric is defined by the horizontal ice flow).
Additionally, the bulk of the sensitive volume probed by ANITA
is slower-moving, thicker interior ice, for which birefringent effects are
expected to be somewhat smaller than near the continental
periphery.

A suite of additional measurements at
a variety of icecap locations, perhaps in association with
future traverses, are needed to fully map out ice properties
across the continent.
In principle, the extensive measurements conducted
to determine the Antarctic and Greenland ice thicknesses can
also be searched for time delays as a function of transmitter/receiver
orientation. Those studies are currently underway.
Particularly important is the determination of 
dielectric properties measured in the vertical plane, for which
there is thus far somewhat less {\it in situ} data.

\section*{Acknowledgments}
ANITA is supported under NASA Grant NAG5-5387.
We also thank the National Science Foundation's
Office of Polar Programs for support
under grant NSF OPP-0338219, the Research Corporation, 
the University of Kansas
Undergraduate Research Award Program, and the NSF's Research
Experience for Undergraduates (REU) program.
We thank the NASA LDB program, and in particular
W. Vernon Jones, for providing the support necessary
for establishment of our field camp.
We thank Phil Austin, RPSC and
Fixed Wing support for their superb logistical assistance.
Jessica Walker provided important surface and bedrock
elevation data while we were in the field. We thank
the members of the ITASE traverse for their assistance in
establishing our field camp, drilling the hole
used for our in-ice transmission, and providing the density
data from the borehole. 
Particulary, we thank Steve Arcone, Daniel Dixon,
Paul Mayewski, and Bryan Welch.
Bryan Welch also provided
essential guidance in assessing the bottom reflection.
Ken Ratzlaff of the KU
IDL lab provided the instrumentation necessary for the
temperature profile measurements. %Inestimable gratitude is owed to Matt Smith and Jessy Jenkins, without whom these measurements would have been, simply, impossible.
DZB wishes to 
thank his RICE colleagues for useful conversations regarding the
birefringence measurement. 

\bibliographystyle{plain}

\begin{thebibliography}{...}
%\bibitem[Price-1997]Optical Properties of Pure Ice at the South Pole: Scattering, 
\bibitem{Buford1997}P. B. Price and L. Bergstrom, Applied Optics {\bf 36}, 4181-4194 (1997).
%\bibitem[Barwick-2005]
\bibitem{Bar2005}S. Barwick, D. Besson, P. Gorham, D. Saltzberg, J. Glac. 04J067 (2005).
%\bibitem[Kravchenko-2006]
\bibitem{Kr2006}I. Kravchenko et al., Phys. Rev. {\bf D73}, 082002 (2006).
%\bibitem[ANSMET]
\bibitem{ANSMET}\begin{verbatim}http://geology.cwru.edu/~ansmet/\end{verbatim}
%\bibitem[Barwick-2006]
\bibitem{Bar2006}S. W. Barwick et al., Phys. Rev. Lett. {\bf 96}, 171101 (2006).
%\bibitem[Jackson-1975]
\bibitem{Jackson75}J. D. Jackson,{\it Classical Electrodynamics}, 2nd Ed., John Wiley \& Sons, New York (1975).
%\bibitem[Hargreaves-1977]  The polarization of radio signals in the  radio echo sounding of ice sheets,; 
\bibitem{Hargreaves}N. D. Hargreaves, J. Phys. D: Applied Physics 10(9), 1285-1304 (1977).
%\bibitem[Hargreaves-1978]; The radio frequency birefringence of polar ice
\bibitem{Hargreaves-1978}N. D. Hargreaves, N.D., J. Glac. 21, 301-313 (1978).
%\bibitem[Matsuoka-2003]``Crystal orientation fabrics within the Antarctic ice sheet revealed by a multipolarization plane and dual-frequency radar survey'',
\bibitem{Matsuoka-biref}K. Matsuoka, T. Furukawa, S. Fujita, N. Maeno N,
  S. Uratsuka, R. Naruse, O. Watanabe, J. Geophys. Res. 108({\bf B10}): 24999, doi:10.1029/2003JB002425 (2003).
%Matsuoka et al., JOURNAL OF GEOPHYSICAL RESEARCH, VOL. 108, NO. B10, 2499, doi:10.1029/2003JB002425 (2003)
%\bibitem[Matsuoka-2004]Ice-flow-induced scattering zone within the Antarctic ice sheet revealed by high-frequency airborne radar,;
\bibitem{Matsuoka2004}K. Matsuoka, S. Uratsuka, S. Fujita, F. Nishio, J. Glac. 50(170), 382-388 (2004).
%\bibitem[Doake-2002]
\bibitem{Doake2002}C. Doake et al., Ann. Glac. {\bf 34}(1), 165-170 (2002).
%\bibitem[Doake-2003]
\bibitem{Doake2003}C. Doake, H. Corr, H. Coor, A. Jenkins, K. Nichols,
  C. Stewart, Euro. Space Agency pub. {\bf 529}, 313-320 (2003);
C. Doake, H. Corr, H. Coor, A. Jenkins, K. Nichols, C. Stewart, FRISP
  Rep. No. 14 (2003).
%Proc. of Workshop on Applications of SAR Polarimetry and Polarimetric Interferometry; Doake CSM, Coor HFJ, Jenkins A, Nicholls KW, and Stewart C ``Interpretation of polarization behaviour of radar waves transmitted through Antarctic ice shelves; Space Agency'', (Special Publication) ESA SP, n 529, p 313-320 (2003); Doake, CSM, Coor HFJ, Jenkins A, Nicholls KW, Stewart C, ``Interpretation of polarimetric ice penetrating radar data over Antarctic ice shelves'' FRISP Report No. 14 (2003).
\bibitem{WoodruffDoake1979}A. H. Woodruff and C. S. M. Doake, J. Glac., 23, 223 (1979)
\bibitem{Fujita-2003}S. Fujita, K. Matsuoka, H. Maeno, T. Furukawa, Ann. Glac 37, 305 (2003).
%\bibitem[Winebrenner-2005]
\bibitem{SIPLE-attenlen}J. D. Winebrenner et al., Ann. Glac., {\bf 37}, 1, 226 (2005)
\bibitem{r:ralston05}J. P. Ralston, Phys. Rev. {\bf D71}, 011503(R) (2005).
%\bibitem[Fitzpatrick-1994]``Preliminary report on the physical and stratigraphic properties of the Taylor Dome ice core.'',
\bibitem{Fitzpatrick-1994}J. J. Fitzpatrick, Antarctic Journal of the U.S. {\bf 29}(5), 84-86 (1994).
%\bibitem[Dixon-2007]
\bibitem{Dixon-2007}D. Dixon, private communication.
%\bibitem[Welch-2006]
\bibitem{Welch-2006}B. Welch, private communication.
%\bibitem[BEDMAP]
\bibitem{BEDMAP}\begin{verbatim}www.antarctica.ac.uk/Resources/AEDC/bedmap/ \end{verbatim}
%\bibitem[Millar-1981]Radio-echo layering in polar ice sheets and past volcanic activity,;
\bibitem{Millar-1981}D. H. M. Millar, Nature, {\bf 292}(5822):441, 443 (1981).
%\bibitem[Bogorodsky-1985]
\bibitem{Bogorodsky}V. V. Bogorodsky, C. R. Bentley, and P. E. Gudmandsen, {\it ed.}, {\it Radioglaciology}.  Dordrecht, Holland, D.Reidel Pub Co., 254 pp. (1985).
%\bibitem[Smith-1972]``Radio echo sounding: absorption and scattering by water inclusion and ice lenses'',
\bibitem{Smith&Evans}B. M. E. Smith. and  S. Evans, J. Glac. {\bf 11}, (61), 133-146 (1972).
%\bibitem[Rippin-2003]Basal topography and ice flow in the Bailey/Slessor region of East  Antarctica,;
\bibitem{Rippin-2003}D. M. Rippin, J. L. Bamber, M. J. Siegert,  J. of Geophys. Res. 108(F1): 6008, doi:10.1029/2003JF000039 (2003).


%\bibitem[Fujita-1999]Nature of radio echo layering in the Antarctic ice sheet detected by a two-frequency experiment,
\bibitem{Fujita-1999}S. Fujita, H. Maeno, S. Uratsuka, T. Furukawa,
  S. Mae, Y. Fujii, and O. Watanabe, J. Geophys. Res. 104({\bf B6}):13013--13023 (1999).
%\bibitem[Matsuoka-1997]
\bibitem{Matsuoka97}T. Matsuoka, S. Fujita, S. Morishima and S. Mae,
  J. Appl. Phys. {\bf 81}(5), 2344 (1997).
%\bibitem[Kravchenko-2005]
\bibitem{Kr2005}I. Kravchenko, D. Besson, and J. Meyers, J. Glac. 03J061 (2005).
%\bibitem[Schytt-1958]``The inner structure of the ice shelf at Maudheim as shown by core drilling.''
\bibitem{Schytt58}V. Schytt, Norwegian-British-Swedish Antarctic Expedition, 1949-52, Scientific Results 4, Glaciology 2., Oslo, Norsk Polarinstitutt. 115-151 (1958).
%\bibitem[Halzen-2006]\bibitem{Francis2006}F. Halzen, Eur. Phys. J. {\bf C46}, 669-687 (2006).
%\bibitem[Fujita-1993]``Relation between ice sheet internal radio-echo reflections and ice fabric at Mizuho Station,  Antarctica'' 
\bibitem{Fujita-1993}S. Fujita and S. Mae, Ann. Glac. {\bf 17}, 269-75 (1993).
%\bibitem[Fujita-2006]
\bibitem{Fujita-2006}S. Fujita et al., J. Glac. {\bf 52}(178), 407 (2006).
%\bibitem[Siegert-2001]``Internal ice-sheet radar layer profiles and their relation to reflection mechanisms between Dome C and the Transantarctic Mountains'' \bibitem{Siegert-2001}M. J. Siegert and S. Fujita, J. Glac. {\bf 47}(157), 205-212 (2001).
%\bibitem[Thorsteinsson-1997]``Textures and fabrics in the GRIP ice core'' \bibitem{Thorsteinsson-1997}T. Thorsteinsson, J. Kipfstuhl and H. Miller,  J. of Geophys. Res. {\bf 102}(C12), 26583-26600, 10.1029/97JC00161 (1997).
\end{thebibliography}

\end{document}